# Matter and Space with Torsion

**Alex Karpelson**

**Abstract:**

General equations of the unified field theory, obtained using the curved and torsional space-time, are presented. They contain only independent geometrical parameters (metric and connections) of the metric-affine space, and describe the distribution and motion of matter, which represents the curvature and torsion of the space-time. Equations, describing spherically symmetric fields, are derived for various cases: gravitational field in vacuum and arbitrary material fields. Equations with and without cosmological terms, are derived for the closed and open models of the uniform and isotropic Universe for the gravitational field in vacuum and arbitrary material fields. The respective solutions without singularities have been obtained for these cases. Equations, describing cosmological models, are presented for various material uniform isotropic fields, but their solutions (formulae for the curvature radius) have been obtained only for the simplest cases – the massless uniform isotropic fields. The relations of the obtained results to the anthropic principle, cosmological natural selection concept, and Gödel theorems are discussed. It is shown that there is no necessity to quantize the obtained general equations.

## 1. Introduction

Our objective is to obtain equations of the unified field theory, based on the 4-dimensional (4D) general metric-affine space with curvature and torsion, and solve these equations for some special cases describing the distribution and motion of matter. We proceed on the basis of the unified geometrical field theory (developed by Einstein, Eddington, Weyl, Schrodinger, Heisenberg, and others), assuming that gravitational field, all physical interactions (weak, electromagnetic and strong) transmitted by bosons, and sources of these interactions, i.e. fermions with half-integer spin (electrons, nucleons, etc.), are merely the manifestations of the unified field. Thus, the elementary particles are just stable local states of "excitation" of the space-time unified geometrical field. In other words, the elementary particles are something like the "geometrodynamic excitons" – the quasi-particles of this theory, i.e. the local concentrations of strength/energy of the space-time unified geometrical field.

Following Clifford, Riemann, Cartan, Einstein, Wheeler, Ivanenko, Hehl, Obukhov, Vladimirov, and others, we assume that this unified field, combining all the interactions and elementary particles, is the curved space-time with torsion and non-symmetric geometrical parameters [1-15].

Such curved and torsional space-time can be represented in a general case by the 4D metric-affine space $G_4$. Some specific cases of this space, the Weyl-Cartan space, Weyl space, Riemann-Cartan space $U_4$, and others, were used in different non-symmetric unified field theories, e.g. in the Einstein-Cartan theory. Concept based on such space, should not be regarded as an alternative to the General Relativity (GR), but rather as its natural extension, obtained by removing the constraint of the symmetry of the connections. As result, it preserves the main advantages of the GR, and at the same time, it solves many GR problems, for example, it prevents the appearance of singularities in the black holes and at the Big Bang/Bounce.

## 2. General equations

The most natural way to obtain general equations, determining the curvature and torsion of the space-time unified geometrical field, represented by metric-affine space $G_4$ and describing the respective distribution and motion of matter, is to use the principle of the least action [16].

The classical Lagrangian $L$ has the following form:

$$L=\sqrt{-g}\left(R+2\Lambda\right),\ g=\left|g_{ik}\right|,\ R=R_{ik}\,g^{ik},\ R_{ik}=\frac{\partial\Gamma_{ik}{}^{l}}{\partial x^{l}}-\frac{\partial\Gamma_{il}{}^{l}}{\partial x^{k}}+\Gamma_{ik}{}^{l}\,\Gamma_{lm}{}^{m}-\Gamma_{il}{}^{m}\,\Gamma_{mk}{}^{l}, \qquad (1)$$

where $R_{ik}$ is the Ricci tensor, $R$ is the scalar curvature, $g^{ik}$ and $g_{ik}$ are respectively the contravariant and covariant metric tensors, $g_{ik}g^{kl}=\delta_i^l$, g is the determinant formed from the $g_{ik}$ quantities, $\Gamma_{kl}{}^{i}$ are the affine connections, $\Lambda\approx10^{-52}\mathrm{m}^{-2}$ is the cosmological constant, all mentioned geometrical parameters are in a general case the non-symmetric ones.

Since $\Lambda$ is small, the presence of this term affects insignificantly all physical fields over not too large regions of the space-time however it leads to appearance of the "cosmological" solutions describing the Universe as a whole.

Lagrangian (1) has many limitations; therefore various more complex Lagrangians have been proposed [6-13], e.g. Lagrangians containing sums of the invariants constructed from quadratic and more complex terms (Poincare gauge field strength). However, each of these novel Lagrangians has significant drawbacks. Einstein [1] asserted that more complex Lagrangians should be analyzed only if there are some rigorous and well-grounded physical causes based on the compelling experimental evidences. But so far, there are no cogent and convincing reasons, observational or theoretical, for such a change in the form of the fundamental equations of the theory that have a profound physical significance.

The field equations in $G_4$ space can be derived from the variational principle for the action $S=\int L\,d^4x$ with Lagrangian (1), where the metric tensor $g^{ik}$ and non-symmetric affine connections $\Gamma_{kl}{}^{i}$ are considered *a priori* the independent variables [1-15]. So now in $G_4$ space we have torsion $S_{jk}{}^{i}$, i.e. anti-symmetric part of the affine connection coefficients:

$$S_{jk}{}^{i} = \frac{1}{2}(\Gamma_{jk}{}^{i} - \Gamma_{kj}{}^{i}) \qquad (2)$$

We also have a contorsion $K_{ijk} = S_{ijk} - S_{jki} + S_{kij}$, i.e. difference between the connection with torsion and the corresponding connection without torsion. Moreover, there is now no metricity condition [1-2, 10, 14, 17], i.e. the non-metricity tensor $Q_{lik}$ (equal to the covariant derivative of the metric tensor $\nabla_l g_{ik}$, measuring compatibility of the independent geometrical parameters $g_{ik}$ and $\Gamma_{ik}{}^{l}$ in G4 space, is now not equal to zero:

$$Q_{lik} = \nabla_l g_{ik} = \frac{\partial g_{ik}}{\partial x^l} - g_{ik}\Gamma^m_{ik} - g_{im}\Gamma^m_{kl} \neq 0 \qquad (3)$$

Thus, we will vary action $S$ by metric $g_{ik}$ and connections $\Gamma_{kl}{}^{i}$ independently (the Palatini principle). Action S should be invariant under the general coordinate transformations combined with the local Lorentz rotations. Recall, that action $S$ of the system with Lagrangian $L$ equals:

$$S = \int L\, d^4 x, \qquad (4)$$

where $d^4 x$ is the 4D volume element.

Variation of action (4) by metric $g_{ik}$ yields

$$R_{ik} - \frac{1}{2} g_{ik} R - \Lambda g_{ik} = 0\,. \qquad (5)$$

Equations (5) are similar to the well-known Einstein equations in the Riemannian space V4 without matter and with cosmological term, but now they are obtained for the metric-affine space G4 with torsion.

Using results from [1-15] and varying action (4) by connections $\Gamma_{kl}{}^{i}$, we obtain

$$\Gamma_{lm}{}^{j} g^{lm} \delta^k{}_i + \Gamma_{il}{}^{l} g^{jk} - \Gamma_{li}{}^{j} g^{lk} - \Gamma_{il}{}^{k} g^{jl} - \frac{1}{\sqrt{-g}} \frac{\partial\left(\sqrt{-g}\, g^{jk}\right)}{\partial x^i} + \frac{\delta^k{}_i}{\sqrt{-g}} \frac{\partial\left(\sqrt{-g}\, g^{jl}\right)}{\partial x^l} = 0, \qquad (6)$$

where $\delta^k{}_i = \begin{cases} 0\,, & i \neq k \\ 1\,, & i = k \end{cases}$ is the Kronecker tensor.

Thus, not assuming *a priori* any symmetry in the connections $\Gamma_{kl}{}^{i}$ and metric $g_{ik}$, and any correlation between $g_{ik}$ and $\Gamma_{kl}{}^{i}$ in G4 space, we obtained the following:

1. The Einstein equations (5) with cosmological term after variation of action (4) by $g_{ik}$. System (5) contains in a general non-symmetric case sixteen partial differential equations with sixteen unknown functions $g_{ik}$ and sixty four unknown functions $\Gamma_{kl}{}^{i}$.
2. Equations (6), connecting the geometrical parameters (metric $g^{ik}$ and connections $\Gamma_{kl}{}^{i}$) of G4 space, as a result of variation of action (4) by $\Gamma_{kl}{}^{i}$. System of equations (6) contains in a general non-symmetric case sixty four partial differential equations with sixty four unknown non-symmetric functions $\Gamma_{kl}^{\ \ i}$ and sixteen unknown functions $g^{ik}$. Equations (6) do not contain terms with cosmological constant $\Lambda$, because the respective cosmological term in Lagrangian (1) does not depend on the connections.

Note that equations (6) are similar to different types of the respective equations, connecting $g^{ik}$ with $\Gamma_{kl}^{\ \ i}$ and obtained in [1-2, 8-14] for various non-symmetric unified field theories.

Emphasize, that Einstein equations of GR, equations of the Einstein-Cartan theory of gravitation with torsion, and equations of various non-symmetric topological gravitational theories and gauge field theories [1-15] are dual. It means, their left-hand sides are related only to the geometry of the space-time, while the right-hand sides are related only to the matter, which represents the external source of the space-time geometry. As result, the matter and geometry are kept separate. This "inconsistency" has been noticed a long time ago by Einstein and others, and they tried to solve this problem.

Unlike all mentioned theories, equations (5)-(6), obtained in $G_4$ space, do not possess such a duality. They contain only the geometrical parameters: metric $g_{ik}$ and non-symmetric connections $\Gamma_{kl}^{\ \ i}$ of $G_4$ space. It means, these equations describe the distribution and motion of matter, representing curvature and torsion of the space-time. Equations (5)-(6) contain in a general case eighty transcendental partial differential equations with eighty unknown functions: sixteen $g_{ik}$ and sixty four $\Gamma_{kl}^{\ \ i}$. These equations are valid for the metric-affine space $G_4$ with torsion and non-symmetric geometrical parameters. Probably, the torsion field represents the spin properties of matter; and may be the spinor theory should be included in general unified field theory for the description of fermions.

Usually, there are some considerations (related to $g_{ik}$ symmetry, $\Gamma_{kl}^{\ \ i}$ symmetry, choice of the reference system, coordinate transformations, and other factors), which allow decreasing the number of equations and unknowns.

Equations (5)-(6) match with the GR equations for a limiting case in the Riemannian $V_4$ space: equations (6) are reduced to the metricity condition $\nabla_l g_{ik} = 0$ (compare with non-metricity tensor (3)), and connections $\Gamma_{kl}^{\ \ i}$ become symmetric $\Gamma_{kl}^{\ \ i} = \Gamma_{lk}^{\ \ i}$ (compare with torsion (2)).

Emphasize that $G_4$ geometry of the space-time arises naturally as a gauge theory for the general affine group [6-15]. For example, some of the gauge gravitational theories, based on the relativity and equivalence principles and associated with the Poincare group (and/or the Lorentz group), lead not to the GR, but to its metric-affine generalization, i.e. to the theory with torsion [12-14]. Note that the multidimensional theories of Kaluza-Klein type can be also interpreted as the geometrical gauge fields [12, 15].

However, all attempts to describe gravitation by using various gauge theories encounter serious problems [7-15]. In general, over the last seven decades many different formulations of gravity with torsion have emerged. Since there are so many different theories, it is natural to assume that most of them are probably wrong.

Also note that, despite its successes in describing the *macroscopic* gravitational phenomena, the Einstein's GR still lacks the status of a fundamental *microscopic* theory, because of the quantization problem and the existence of singular solutions under very general assumptions. Among many attempts to overcome these difficulties, various multidimensional theories and gauge theories of gravity are especially attractive, as the concept of the gauge symmetry has been very successful in the foundation of other fundamental interactions.

## 3. Spherically symmetric fields

### 3.1. Field equations

Analyze equations (5)-(6) for spherically symmetric non-stationary field in $G_4$ space. In such a case, in the 4D spherical coordinates $x^0 = ct$, $x^1 = r$, $x^2 = \theta$, $x^3 = \varphi$ (where c is the speed of light), we have only diagonal non-zero components of metric tensor [16]:

$$g_{00} = \exp\left[\nu(r,t)\right],\ g_{11} = -\exp\left[\lambda(r,t)\right],\ g_{22} = -r^2,\ g_{33} = -r^2\sin^2\theta,\ g_{ik}g^{kl} = \delta_i^l,$$

$$g^{00} = \exp\left[-\nu(r,t)\right],\ g^{11} = -\exp\left[-\lambda(r,t)\right],\ g^{22} = -r^{-2},\ g^{33} = -r^{-2}\sin^{-2}\theta, \tag{7}$$

where $\nu(r,t)$ and $\lambda(r,t)$ are the unknown functions describing the space-time geometry.

As one can see from (7), in the case of spherically symmetric field, the metric components $g_{00}$, $g_{11}$, $g^{00}$, and $g^{11}$, depending on the unknown functions $\nu(r,t)$ and $\lambda(r,t)$, may have singularities; and these singularities will be the real physical singularities, not the fictitious ones which can be removed by the coordinate transformation.

Due to spherical symmetry of the analyzed field the number of equations in (5)-(6) and number of unknowns are much less than the respective numbers in a general non-symmetric case.

Substituting (7) in (6), one obtains the following expressions for connections $\Gamma_{jk}{}^{i}$

$$\Gamma_{01}{}^{0} = \frac{\nu' - \lambda'}{2} + \Gamma_{11}{}^{1},\ \Gamma_{02}{}^{0} = \Gamma_{22}{}^{2},\ \ \Gamma_{03}{}^{0} = \Gamma_{33}{}^{3},\ \Gamma_{10}{}^{0} = \frac{\nu'}{2},\ \Gamma_{11}{}^{0} = \frac{c}{4}\exp(\lambda - \nu)\left(\dot{\lambda} - \dot{\nu}\right),$$

$$\Gamma_{00}{}^{1} = \frac{\nu'}{2}\exp(\nu - \lambda),\ \Gamma_{01}{}^{1} = \frac{c}{4}\left(\dot{\lambda} - \dot{\nu}\right),\ \Gamma_{10}{}^{1} = \Gamma_{00}{}^{0} + \frac{c}{2}\left(\dot{\lambda} - \dot{\nu}\right),\ \Gamma_{22}{}^{1} = -r\exp(-\lambda),$$

$$\Gamma_{12}{}^{1} = \Gamma_{22}{}^{2},\ \Gamma_{33}{}^{1} = -r\sin^2\theta\exp(-\lambda),\ \Gamma_{13}{}^{1} = \Gamma_{33}{}^{3},\ \ \Gamma_{21}{}^{2} = \frac{1}{r} - \frac{\lambda'}{2} + \Gamma_{11}{}^{1},\ \Gamma_{12}{}^{2} = \frac{1}{r}, \tag{8}$$

$$\Gamma_{33}{}^{2} = -\sin\theta\cos\theta,\ \Gamma_{20}{}^{2} = \Gamma_{00}{}^{0} + \frac{c}{4}\left(\dot{\lambda} - \dot{\nu}\right), \Gamma_{23}{}^{2} = \Gamma_{33}{}^{3},\ \ \Gamma_{13}{}^{3} = \frac{1}{r},\ \Gamma_{23}{}^{3} = \cot\theta,$$

$$\Gamma_{31}{}^{3} = \frac{1}{r} - \frac{\lambda'}{2} + \Gamma_{11}{}^{1},\ \Gamma_{32}{}^{3} = \cot\theta + \Gamma_{22}{}^{2},\ \Gamma_{30}{}^{3} = \Gamma_{00}{}^{0} + \frac{c}{4}\left(\dot{\lambda} - \dot{\nu}\right),$$

where four connections $\Gamma_{00}^{\ \ 0}$, $\Gamma_{11}^{\ \ 1}$, $\Gamma_{22}^{\ \ 2}$, $\Gamma_{33}^{\ \ 3}$ are undetermined, all other $\Gamma_{jk}^{\ \ i}$ are equal to zero, the prime and the dot on a symbol denote differentiation with respect to $r$ and t respectively.

Four undetermined quantities $\Gamma_{00}^{\ \ 0}$, $\Gamma_{11}^{\ \ 1}$, $\Gamma_{22}^{\ \ 2}$, $\Gamma_{33}^{\ \ 3}$ appear in solution (8), because number of the independent equations in system (5)-(6) for spherically symmetric field, unlike any general non-symmetric field, is less than number of the variables. Probably, this indefiniteness is similar to the indefiniteness mentioned in [18], where in the presence of curvature and torsion, the space of the Lorentz connections becomes the affine space, and consequently one can always add a tensor to a given connection without destroying the covariance of the theory.

From (5), (7), and (8) one can obtain equations for the spherically symmetric field in G4 space:

$$R_{00}-\frac{1}{2}g_{00}R=\left(\frac{\lambda'}{r}-\frac{1}{r^2}\right)\exp(\nu-\lambda)+\frac{\exp(\nu)}{r^2}=0 \quad ,$$

$$R_{11}-\frac{1}{2}g_{11}R=\left(\frac{\nu'}{r}+\frac{1}{r^2}\right)-\frac{\exp(\lambda)}{r^2}=0 \, ,$$

$$R_{22}-\frac{1}{2}g_{22}R=\left[r\nu''+\frac{r(\nu')^2}{2}-\frac{r\nu'\lambda'}{2}-\lambda'+\nu'\right]\frac{r\exp(-\lambda)}{2}-\frac{c^2r^2}{4}\exp(-\nu)\left[\frac{(\dot\lambda-\dot\nu)^2}{2}+(\ddot\lambda-\ddot\nu)\right]=0$$

$$\begin{aligned} R_{33}-\frac{1}{2}g_{33}R=&\left[r\nu''+\frac{r(\nu')^2}{2}-\frac{r\nu'\lambda'}{2}-\lambda'+\nu'\right]\frac{r\exp(-\lambda)\sin^2\theta}{2}- \\ &-\frac{c^2}{4}r^2\sin^2\theta\exp(-\nu)\left[\frac{(\dot\lambda-\dot\nu)^2}{2}+(\ddot\lambda-\ddot\nu)\right]=0 \end{aligned} \quad ,$$

$$R_{01}-\frac{1}{2}g_{01}R=\frac{c}{2}\frac{\partial}{\partial t}(\nu'-\lambda')+\frac{c}{2r}(\dot\lambda-\dot\nu)+c\frac{\partial\Gamma_{11}^{\ \ 1}}{\partial t}-\frac{\partial\Gamma_{00}^{\ \ 0}}{\partial r}=0 \; , \qquad (9)$$

$$R_{10}-\frac{1}{2}g_{10}R=-\frac{c}{2}\frac{\partial}{\partial r}(\dot\nu-\dot\lambda)+\frac{c}{2r}(\dot\lambda-\dot\nu)-c\frac{\partial\Gamma_{11}^{\ \ 1}}{\partial t}+\frac{\partial\Gamma_{00}^{\ \ 0}}{\partial r}=0 \quad ,$$

$$R_{02}-\frac{1}{2}g_{02}R\equiv R_{02}=-R_{20}+\frac{1}{2}g_{20}R\equiv -R_{20}=c\frac{\partial\Gamma_{22}^{\ \ 2}}{\partial t}-\frac{\partial\Gamma_{00}^{\ \ 0}}{\partial\theta}=0 \; ,$$

$$R_{03}-\frac{1}{2}g_{03}R\equiv R_{03}=-R_{30}+\frac{1}{2}g_{30}R\equiv -R_{30}=c\frac{\partial\Gamma_{33}^{\ \ 3}}{\partial t}-\frac{\partial\Gamma_{00}^{\ \ 0}}{\partial\varphi}=0 \quad ,$$

$$R_{12}-\frac{1}{2}g_{12}R\equiv R_{12}=-R_{21}+\frac{1}{2}g_{21}R\equiv -R_{21}=\frac{\partial\Gamma_{22}^{\ \ 2}}{\partial r}-\frac{\partial\Gamma_{11}^{\ \ 1}}{\partial\theta}=0 \quad ,$$

$$R_{13}-\frac{1}{2}g_{13}R\equiv R_{13}=-R_{31}+\frac{1}{2}g_{31}R\equiv -R_{31}=\frac{\partial\Gamma_{33}^{\ \ 3}}{\partial r}-\frac{\partial\Gamma_{11}^{\ \ 1}}{\partial\varphi}=0 \quad ,$$

$$R_{23}-\frac{1}{2}g_{23}R\equiv R_{23}=-R_{32}+\frac{1}{2}g_{32}R\equiv -R_{32}=\frac{\partial\,\Gamma_{33}{}^{3}}{\partial\,\theta}-\frac{\partial\,\Gamma_{22}{}^{2}}{\partial\,\varphi}=0\quad, \tag{9}$$

where the double prime on a symbol denotes double differentiation with respect to $r$, and the double dot on a symbol denotes double differentiation with respect to t.

Note that the fourth equation in (9) can be easily reduced to the third one; it means the forth equation is not independent, and therefore can be disregarded.

System (9) was obtained from equations (5)-(6) at the condition that cosmological constant $\Lambda$ equals to zero, because the "cosmological term" $\Lambda g_{ik}$ in (5) is related, probably, to the "vacuum space-time", since vacuum has non-zero energy and non-zero mass. However, the "vacuum mass" is significant only on the cosmological scale; so the term $\Lambda g_{ik}$ in (5) can be neglected on the smaller scale of the spherically symmetric field.

Now analyze system (9) for two different cases:

1. There are no singularities in functions $\nu(r,t)$ and $\lambda(r,t)$. But if $\nu(r,t)$ and $\lambda(r,t)$ are continuous functions, then, because of the Clairaut theorem based on the continuity condition, their second-order successive mixed partial derivatives with respect to different variables are equal:

$$\frac{\partial^2\lambda(r,t)}{\partial r\,\partial t}=\frac{\partial^2\lambda(r,t)}{\partial t\,\partial r}\quad\text{and}\quad\frac{\partial^2\nu(r,t)}{\partial r\,\partial t}=\frac{\partial^2\nu(r,t)}{\partial t\,\partial r}. \tag{10}$$

2. There are singularities in functions $\nu(r,t)$ and $\lambda(r,t)$. It means that $\nu(r,t)$ and $\lambda(r,t)$ are now the discontinuous functions, and therefore their second-order successive mixed partial derivatives with respect to different variables are not equal:

$$\frac{\partial^2\lambda(r,t)}{\partial r\,\partial t}\neq\frac{\partial^2\lambda(r,t)}{\partial t\,\partial r}\quad\text{and}\quad\frac{\partial^2\nu(r,t)}{\partial r\,\partial t}\neq\frac{\partial^2\nu(r,t)}{\partial t\,\partial r}. \tag{11}$$

### 3.2. Gravitational fields in vacuum with functions $\nu(r,t)$ and $\lambda(r,t)$ without singularities

In the case, when there are no singularities in functions $\nu(r,t)$ and $\lambda(r,t)$, the first six equations in system (9) containing these functions and their derivatives, can be significantly simplified. It is also necessary to make an assumption regarding the undetermined connections $\Gamma_{00}{}^{0}$, $\Gamma_{11}{}^{1}$, $\Gamma_{22}{}^{2}$, $\Gamma_{33}{}^{3}$. To obtain from (8) and (9) the simplest of all solutions (i.e. solution related to the minimal curvature and torsion of the space-time), we assume that

$$\Gamma_{00}{}^{0}=\Gamma_{11}{}^{1}=\Gamma_{22}{}^{2}=\Gamma_{33}{}^{3}=0 \tag{12}$$

Substitution of (12) into (8) gives

$$\Gamma_{01}{}^{0} = \frac{\nu' - \lambda'}{2},\ \Gamma_{10}{}^{0} = \frac{\nu'}{2},\ \Gamma_{00}{}^{1} = \frac{\nu'}{2}\exp(\nu - \lambda),\ \Gamma_{11}{}^{0} = \frac{c}{4}\exp(\lambda - \nu)\left(\dot{\lambda} - \dot{\nu}\right),$$

$$\Gamma_{01}{}^{1} = \frac{c}{4}\left(\dot{\lambda} - \dot{\nu}\right),\quad \Gamma_{10}{}^{1} = \frac{c}{2}\left(\dot{\lambda} - \dot{\nu}\right),\ \Gamma_{20}{}^{2} = \frac{c}{4}\left(\dot{\lambda} - \dot{\nu}\right),\ \Gamma_{30}{}^{3} = \frac{c}{4}\left(\dot{\lambda} - \dot{\nu}\right), \tag{13}$$

$$\Gamma_{22}{}^{1} = -r\exp(-\lambda),\ \Gamma_{33}{}^{1} = -r\sin^{2}\theta\exp(-\lambda),\ \Gamma_{12}{}^{2} = \Gamma_{13}{}^{3} = \frac{1}{r},\ \Gamma_{21}{}^{2} = \Gamma_{31}{}^{3} = \frac{1}{r} - \frac{\lambda'}{2},$$

$$\Gamma_{33}{}^{2} = -\sin\theta\cos\theta,\ \Gamma_{23}{}^{2} = \Gamma_{32}{}^{3} = \cot\theta,\quad \text{all other } \Gamma_{jk}{}^{i} \text{ are equal to zero.}$$

Assumption (12), leading to the solution (13), can be explained as follows. The different sets of connections $\Gamma_{jk}{}^{i}$ are related to the various degrees of curvature and torsion of the space-time. Formulae (13) represent the simplest set of $\Gamma_{jk}{}^{i}$, because any other set of connections $\Gamma_{jk}{}^{i}$, which is not based on the condition (12), leads to a more complex solution. Therefore, system (12)-(13) is the simplest one, and it corresponds to the minimal curvature and torsion of the space-time, describing the gravitational spherically symmetric field in vacuum, i.e. the pure gravitational field outside the masses producing this field. Any other field creates also a gravitational field. This means, that curvature and torsion of the space-time, related to such more complex "mixed" field, and the respective set of connections $\Gamma_{jk}{}^{i}$, must be also more complex. Recall that a gravitational field is the unique one among various physical fields, because it does not lead to the appearance of any other fields. In this sense, it is the "simplest" physical field, and it corresponds to the space-time with minimum curvature and torsion.

Substituting (13) in (9) and keeping in mind that $\nu$ and $\lambda$ are continuous functions, we obtain

$$\begin{gathered}
\left[r\lambda' - 1\right]\exp(-\lambda) + 1 = 0 \\
\left[r\nu' + 1\right]\exp(-\lambda) - 1 = 0 \\
r\nu'' + \frac{r}{2}(\nu')^{2} - \frac{r}{2}\nu'\lambda' - \lambda' + \nu' = 0 \\
\dot{\lambda} = \dot{\nu}
\end{gathered} \tag{14}$$

The third equation in (14) is not an independent one; it can be derived from the first and second equations. The forth equation in (14) was obtained after adding the fifth and sixth equations in (9). Note, that adding the first and second equations in (14), we obtain $\lambda' + \nu' = 0$. Function $\lambda(r)$ can be obtained from the first equation in (14); then function $\nu(r)$ can be determined from the second equation in (14).

Equations (14) coincide with the Schwarzschild equations for the spherically symmetric field in vacuum (i.e. gravitational field without matter) in the Riemannian space $V_4$ [16]. It is not surprising, since both systems of equations were obtained for the pure gravitational spherically symmetric field.

Based on these considerations, the stationary solution of equations (14) can be presented as

$$\lambda = \ln\left(\frac{r}{r+C_1}\right), \quad \nu = \ln\left(C_2\frac{r+C_1}{r}\right), \quad \text{respectively}$$

$$\exp(\nu) = g_{00} = \frac{1}{g^{00}} = C_2\frac{r+C_1}{r}, \quad \exp(\lambda) = -g_{11} = -\frac{1}{g^{11}} = \frac{r}{r+C_1}, \qquad (15)$$

where $C_1$ and $C_2$ are the constants which can be calculated using the boundary conditions.

Further we will analyze only metric tensor components $g_{00}$ and $g_{11}$ because they determine the main parameters of the centrally symmetric field in spherical coordinates.

Formulae (15) show that spherically symmetric gravitational field in vacuum is automatically the stationary one. It just confirms the Birkhoff's theorem, which states that any spherically symmetric solution of the field equations in vacuum must be stationary and asymptotically flat. Physically it means that, first of all, this field should vanish at the large distances and, secondly, if the spherically symmetric gravitational field is not stationary, then it will collapse with time in one center point. Note that the Birkhoff's theorem can be generalized on a spherically symmetric solution of the material field.

Solution (15) in $G_4$ space corresponds to the pure gravitational spherically symmetric stationary field (i.e. gravitational field in vacuum). Such a field can exist either at the large $r$ values outside of the masses producing this field, or at the small $r$ values in the interior of a spherical cavity within a centrally symmetric distribution of masses producing this field. In the first case, we have an extremely dispersed field, i.e. a weak stationary gravitational field at the large distances from the bodies that produce it. Consequently, the space-time, related to such field, must be maximally uniform and simple. In the second case, we obtain gravitational field within a cavity, inside the spherically symmetric distribution of bodies producing this field.

Recall, that in order to analyze the spherically symmetric distribution of matter, we cannot use solution (15), we should obtain an appropriate solution of equations (9), where we cannot use condition (12) anymore, because it is valid only for the pure gravitational field in vacuum.

Now return to formulae (15), describing gravitational spherically symmetric stationary field in vacuum in $G_4$ space. To obtain from (15) a solution for the large $r$, one can use the ordinary boundary conditions: the Newton's expressions for metric at $r=\infty$ [16]. As result, we will get the following constants: $C_1=-r_g$ (where $r_g$ is the gravitational radius) and $C_2=1$. It yields solution (15) coinciding with the Schwarzschild solution for the centrally symmetric stationary gravitational field in spherical coordinates in vacuum away from its material sources:

$$g_{00} = \exp(\nu) = \frac{r-r_g}{r} \quad \text{and} \quad g_{11} = -\exp(\lambda) = -\frac{r}{r-r_g} \qquad (16)$$

Note that solution (15) at very large r coincides, as it should be expected, with the Galilean metric components in spherical coordinates

$$g_{00}=g^{00}=1,\ g_{11}=g^{11}=-1 \qquad (17)$$

Formulae (17) describe a flat space-time without curvature and torsion at the large distances from the masses producing this field, i.e. even a gravitational field does not exist in this area.

To obtain a physically reasonable solution at the small $r$ in the interior of a spherical cavity within the centrally symmetric distribution of the gravitating masses, one should use formulae (15) and the boundary conditions, which exclude singularities of the metric components at $r=0$. The absence of singularities is natural for any closed physical theory. To satisfy it, we should assume that in (15) coefficient $C_1=0$ and coefficient $C_2$ =1. Then we obtain the Galilean metric (17) within this spherical cavity. This result means that there is just a flat space-time without any curvature and torsion inside such spherical cavity. However, a flat space-time means that even a gravitational field does not exist in this area. This is physically reasonable, because even in the Newton's theory there is no gravitational field inside any spherically symmetric cavity within the centrally symmetric distribution of masses.

Thus in general, we can conclude that metric, obtained in $G_4$ space, unlike the respective solution of the Einstein equations in the Riemannian space $V_4$ in the GR, gives no singularities for spherically symmetric stationary gravitational field at r=0 and $r=r_g$. It means that such disputable phenomena as the gravitational collapse and black hole formation do not exist for this field in $G_4$ space. Probably it occurs because, unlike the GR, the torsion violates the energy conditions of the Penrose-Hawking theorems; and therefore the singularities may be avoided. It is also possible, because of the torsion, to obtain solutions without singularities in the Einstein-Cartan theory in $U_4$ space [10].

Generically, the results obtained in sections 3.1 and 3.2, endorse the concept that metric-affine space $G_4$ with curvature and torsion describes the distribution and motion of matter. These results also confirm logic of general formulae (5)-(6), correctness of equations (9) and (14) for the spherically symmetric field, and accuracy of the respective solution (15), because this solution is the stationary one, does not contain singularities, and in the limit coincides with the Schwarzschild solution. Obtained results also back the validity of the assumption (12) that the simplest set (13) of connections $\Gamma_{jk}^{\ \ i}$ is related to the physical field with minimal curvature and torsion, i.e. to the spherically symmetric stationary gravitational field in vacuum.

### 3.3. Material field in space with functions $\nu(r,t)$ and $\lambda(r,t)$ without singularities

Now let us try to solve system (9), describing not a gravitational spherically symmetric field in vacuum, but an arbitrary material spherically symmetric field with functions $\nu(r,t)$ and $\lambda(r,t)$ without singularities. In case of a material field, we cannot use condition (12), which is valid only for the gravitational field in vacuum. However, system (9) can be simplified as follows.

First of all, simplify the first two equations in (9), then note that the third and fourth equations in (9) are similar (as it has already been mentioned above), after that add the fifth equation in (9) to the sixth equation, and finally subtract the fifth equation in (9) from the sixth equation.

As result, the complex system (9) will be reduced to the following rather simple system of equations:

$$
\begin{gathered}
[r\lambda'-1]\exp(-\lambda)+1=0\ , \\
[r\nu'+1]\exp(-\lambda)-1=0\ , \\
\left[\nu''+\frac{(\nu')^2}{2}-\frac{\nu'\lambda'}{2}+\frac{\nu'-\lambda'}{r}\right]-\frac{c^2}{2}\exp(\lambda-\nu)\left[\frac{(\dot{\lambda}-\dot{\nu})^2}{2}+(\ddot{\lambda}-\ddot{\nu})\right]=0\,, \\
\dot{\lambda}=\dot{\nu}\,, \\
c\frac{\partial^2(\nu-\lambda)}{\partial t\,\partial r}+2c\frac{\partial\Gamma_{11}{}^{1}}{\partial t}-2\frac{\partial\Gamma_{00}{}^{0}}{\partial r}=0\,, \\
c\frac{\partial\Gamma_{22}{}^{2}}{\partial t}-\frac{\partial\Gamma_{00}{}^{0}}{\partial\theta}=0\,, \qquad\qquad (18) \\
c\frac{\partial\Gamma_{33}{}^{3}}{\partial t}-\frac{\partial\Gamma_{00}{}^{0}}{\partial\varphi}=0\,, \\
\frac{\partial\Gamma_{22}{}^{2}}{\partial r}-\frac{\partial\Gamma_{11}{}^{1}}{\partial\theta}=0\,, \\
\frac{\partial\Gamma_{33}{}^{3}}{\partial r}-\frac{\partial\Gamma_{11}{}^{1}}{\partial\varphi}=0\,, \\
\frac{\partial\Gamma_{33}{}^{3}}{\partial\theta}-\frac{\partial\Gamma_{22}{}^{2}}{\partial\varphi}=0
\end{gathered}
$$

If we substitute the forth equation in (18) in the third equation, then the first four equations in system (18), describing a material spherically symmetric field, will match with equations (14), describing the gravitational spherically symmetric field in vacuum. Then the left term in the fifth equation, presenting the second-order successive mixed derivative, becomes equal to zero.

Thus, the first four equations in (18) (or equations (14)) determine functions $\nu(r,t)$ and $\lambda(r,t)$ while the last six equations in (18) determine four connections $\Gamma_{00}{}^{0}$, $\Gamma_{11}{}^{1}$, $\Gamma_{22}{}^{2}$, $\Gamma_{33}{}^{3}$.

Respectively, the obtained results for functions $\nu(r,t)$ and $\lambda(r,t)$ and metric components $g_{00}$ and $g_{11}$ will coincide with solution (15), presented above and describing the stationary spherically symmetric pure gravitational field with metric without singularities. It means that metric of the spherically symmetric arbitrary material field can be determined using only the first four equations in (18), i.e. this metric depends only on the gravitational field and does not depend on the matter. In other words, the spherically symmetric arbitrary material field will be stationary and asymptotically flat, which confirms the generalized Birkhoff's theorem.

However recall, that in order to determine the whole geometry of a material field in spherically symmetric metric-affine space $G_4$ with curvature and torsion, we need all $\Gamma^{i}_{jk}$ from (8) and four unknown $\Gamma_{00}{}^{0}$, $\Gamma_{11}{}^{1}$, $\Gamma_{22}{}^{2}$, $\Gamma_{33}{}^{3}$, which should be determined from the last six equations in (18).

Using considerations, mentioned above and concerning system (14), equations (18) can be simplified, and the final system of equations, describing spherically symmetric material field in $G_4$ space with functions $\nu(r,t)$ and $\lambda(r,t)$ without singularities, can be presented as follows:

$$
\begin{aligned}
& r\lambda' - 1 + \exp(\lambda) = 0 \, , \\
& r\nu' + 1 - \exp(\lambda) = 0 \, , \\
& c\frac{\partial \Gamma_{11}{}^{1}}{\partial t} - \frac{\partial \Gamma_{00}{}^{0}}{\partial r} = 0 \, , \\
& c\frac{\partial \Gamma_{22}{}^{2}}{\partial t} - \frac{\partial \Gamma_{00}{}^{0}}{\partial \theta} = 0 \, , \\
& c\frac{\partial \Gamma_{33}{}^{3}}{\partial t} - \frac{\partial \Gamma_{00}{}^{0}}{\partial \varphi} = 0 \, , \\
& \frac{\partial \Gamma_{22}{}^{2}}{\partial r} - \frac{\partial \Gamma_{11}{}^{1}}{\partial \theta} = 0 \, , \\
& \frac{\partial \Gamma_{33}{}^{3}}{\partial r} - \frac{\partial \Gamma_{11}{}^{1}}{\partial \varphi} = 0 \, , \\
& \frac{\partial \Gamma_{33}{}^{3}}{\partial \theta} - \frac{\partial \Gamma_{22}{}^{2}}{\partial \varphi} = 0
\end{aligned}
\tag{19}
$$

Emphasize, that using (19), four connections $\Gamma_{00}{}^{0}$, $\Gamma_{11}{}^{1}$, $\Gamma_{22}{}^{2}$, $\Gamma_{33}{}^{3}$ cannot be determined uniquely. It means these four connections may have different forms, i.e. they may be related to the different spherically symmetric material fields. Formulae (19) and (8) describe the general case – the spherically symmetric arbitrary material field, while formulae (12), (13) and (14), which can be derived from (19) and (8) as a particular case, describe the respective particular case – spherically symmetric pure gravitational field.

In order to determine from (9) connections $\Gamma_{00}{}^{0}$, $\Gamma_{11}{}^{1}$, $\Gamma_{22}{}^{2}$, $\Gamma_{33}{}^{3}$, representing some concrete material spherically symmetric stationary field, the technique described in [19], where it has been applied to the massless Weyssenhoff fluid with spin [20], can be used.

**3.4. Field in space with functions $\nu(r,t)$ and $\lambda(r,t)$ containing singularities**

Now analyze a general case, when there are singularities in functions $\nu(r,t)$ and $\lambda(r,t)$. First of all, using general system (9), we should obtain equations describing the spherically symmetric field with functions $\nu(r,t)$ and $\lambda(r,t)$ containing singularities. To do it, add the fifth equation in (9) to the sixth equation, subtract the fifth equation in (9) from the sixth equation, and also note (as it has already been mentioned) that the third and fourth equations in (9) are equivalent to each other.

Then, instead of (9), we have

$$
\begin{gathered}
r\lambda' - 1 + \exp\lambda = 0 \ , \\
r\nu' + 1 - \exp\lambda = 0 \ , \\
\left[\nu'' + \frac{(\nu')^2}{2} - \frac{\nu'\lambda'}{2} + \frac{\nu' - \lambda'}{r}\right] - \frac{c^2}{2}\exp(\lambda - \nu)\left[\frac{(\dot{\lambda} - \dot{\nu})^2}{2} + (\ddot{\lambda} - \ddot{\nu})\right] = 0 \ , \\
\frac{\partial}{\partial t}(\nu' - \lambda') - \frac{\partial}{\partial r}(\dot{\nu} - \dot{\lambda}) + \frac{2}{r}(\dot{\lambda} - \dot{\nu}) = 0 \ , \\
\frac{c}{2}\frac{\partial}{\partial t}(\nu' - \lambda') + \frac{c}{2}\frac{\partial}{\partial r}(\dot{\nu} - \dot{\lambda}) + 2c\frac{\partial \Gamma_{11}^{\ 1}}{\partial t} - 2\frac{\partial \Gamma_{00}^{\ 0}}{\partial r} = 0 \ , \\
c\frac{\partial \Gamma_{22}^{\ 2}}{\partial t} - \frac{\partial \Gamma_{00}^{\ 0}}{\partial \theta} = 0 \ , \\
c\frac{\partial \Gamma_{33}^{\ 3}}{\partial t} - \frac{\partial \Gamma_{00}^{\ 0}}{\partial \varphi} = 0 \ , \\
\frac{\partial \Gamma_{22}^{\ 2}}{\partial r} - \frac{\partial \Gamma_{11}^{\ 1}}{\partial \theta} = 0 \ , \\
\frac{\partial \Gamma_{33}^{\ 3}}{\partial r} - \frac{\partial \Gamma_{11}^{\ 1}}{\partial \varphi} = 0 \ , \\
\frac{\partial \Gamma_{33}^{\ 3}}{\partial \theta} - \frac{\partial \Gamma_{22}^{\ 2}}{\partial \varphi} = 0 \ .
\end{gathered}
\tag{20}
$$

In a general case, a spherically symmetric non-stationary field cannot be *a priori* considered a field with functions $\nu(r,t)$ and $\lambda(r,t)$ without singularities. Therefore, the second-order successive mixed partial derivatives of these functions are not equal to each other, see (11). In other words, the Clairaut theorem based on the continuity condition is not valid. Respectively, system (20) should be used to describe the spherically symmetric non-stationary field in a general case with singular functions $\nu(r,t)$ and $\lambda(r,t)$. However, this system consists of the complex equations; therefore it is hard to obtain its general solution. Nevertheless, the following technique can be used to solve this system of equations.

**3.5. Gravitational field in vacuum in space with functions $\nu(r,t)$ and $\lambda(r,t)$ containing singularities**

First of all, let us use condition (12) and analyze only the simplest pure gravitational field.

Start from introducing a new function in the fourth equation in system (20):

$$f(r,t)=\left(\dot{\lambda}-\dot{\nu}\right)=\frac{r}{2}\left[\frac{\partial}{\partial r}\left(\dot{\nu}-\dot{\lambda}\right)-\frac{\partial}{\partial t}\left(\nu'-\lambda'\right)\right] \quad . \tag{21}$$

Then substitute (21) and (12) in (20) and obtain the following system of equations:

$$\begin{gathered}
r\lambda'-1+\exp\lambda=0 \ , \\
r\nu'+1-\exp\lambda=0 \ , \\
\left[\nu''+\frac{(\nu')^2}{2}-\frac{\nu'\lambda'}{2}+\frac{\nu'-\lambda'}{r}\right]=\frac{c^2}{2}\exp(\lambda-\nu)\left[\frac{f(r,t)^2}{2}+\frac{\partial f(r,t)}{\partial t}\right] , \\
f(r,t)=\left(\dot{\lambda}-\dot{\nu}\right)=\frac{r}{2}\left[\frac{\partial}{\partial r}\left(\dot{\nu}-\dot{\lambda}\right)-\frac{\partial}{\partial t}\left(\nu'-\lambda'\right)\right] , \\
\frac{\partial}{\partial t}\left(\nu'-\lambda'\right)+\frac{\partial}{\partial r}\left(\dot{\nu}-\dot{\lambda}\right)=0 \quad ,
\end{gathered} \tag{22}$$

System (22) is much simpler than system (20). Using the fourth and fifth equations in (22), it is easy to get the equation for function f(r, t):

$$f(r,t)=-r\frac{\partial f(r,t)}{\partial r} \tag{23}$$

Taking into account that $\int\frac{dx}{x}=\ln|x|$ (where and below symbol |f| means the modulus of function f), equation (23) can be easily solved, and then a general solution of equations (22) can be presented as follows

$$\begin{gathered}
|f(r,t)|=\frac{|A(t)|}{r}, \ \ \nu'=\frac{1}{2r^2}\int A(t)dt\ -\frac{1}{2}B'(r), \ \ \lambda'=\frac{-1}{2r^2}\int A(t)dt\ +\frac{1}{2}B'(r), \\
\dot{\nu}=-\frac{A(t)}{2r}+\frac{1}{2}\dot{C}(t), \ \ \dot{\lambda}=\frac{A(t)}{2r}+\frac{1}{2}\dot{C}(t), \ \ \nu'-\lambda'=\frac{1}{r^2}\int A(t)dt\ -B'(r), \\
\nu=\frac{-1}{2r}\int A(t)dt\ -\frac{1}{2}B(r)+\frac{1}{2}C(t), \ \ \lambda=\frac{1}{2r}\int A(t)dt\ +\frac{1}{2}B(r)+\frac{1}{2}C(t)
\end{gathered} \tag{24}$$

where A(t), B(r) and C(r) are the unknown functions.

General solution (24) satisfies the first, second, fourth, and fifth equations in (22). However, all functions in (24) have singularities at least at r=0. Functions A(t), B(r) and C(r) can be determined by using the third equation in (22) and the respective boundary/initial conditions. Substituting (24) in the third equation in (22), we can present the obtained equation as follows:

$$-\frac{B''(r)}{2}+\frac{[B'(r)]^2}{4}-\frac{B'(r)}{r}=-\frac{1}{4r^2}\left(\int A(t)dt\right)^2+\frac{B'(r)}{2r^2}\int A(t)dt+$$
$$+\frac{c^2}{2}\left[\frac{A^2(t)}{2r^2}+\frac{\dot{A}(t)}{r}\right]\exp\left[\frac{1}{r}\int A(t)dt+B(r)\right] \qquad (25)$$

Note, that equation (25) does not contain function C(t), which is the part of the general solution (24). It confirms statement in [16] that due to the form (7) of metric tensor, we still have the possibility of the arbitrary transformation of the time in the form t=f(t′). Such a transformation is equivalent to adding to ν an arbitrary function of the time; and with its aid we can always make C(t) vanish. And so, without any loss of generality, we can set

$$C(t)=0 \qquad (26)$$

Recall, that deriving equation (25), we got condition (26) automatically. Note also, that substituting (26) in (24), we obtain λ = -ν.

Equation (25), containing two unknown functions A(t) and B(r), should be valid at any r and t. At the same time, the left-hand side of this equation depends only on the variable r and does not depend on the variable t, while the right-hand side depends on both variables: r and t. This can be true only if the right-hand side of this equation does not depend on t. This is possible only if A(t)=0. At this condition, (25) reduces to the following equation for B(r):

$$-\frac{B''(r)}{2}+\frac{[B'(r)]^2}{4}-\frac{B'(r)}{r}=0 \qquad (27)$$

Equation (27) is a general Riccati equation for function B′(r). Its general solution can be easily obtained applying an integration method used for solving general Riccati equations:

$$B'(r)=\frac{2c_1}{r(r+c_1)} \qquad (28)$$

where $c_1$ is the new constant.

Then, respectively, the solution of equation (25) equals

$$A(t)=0 \text{ and } B(r)=2\ln\left|\left(\frac{c_2 r}{r+c_1}\right)\right| , \qquad (29)$$

where $c_1$ and $c_2$ are the constants.

Using (24), (26), (28), and (29), the final solution of system (22) and the respective metric components (7) can be presented as

$$|f(r,t)|=\frac{0}{r}, \quad \nu=\frac{-0}{2r}-\ln\left|\left(\frac{c_2 r}{c_1+r}\right)\right|, \quad \lambda=\frac{0}{2r}+\ln\left|\left(\frac{c_2 r}{c_1+r}\right)\right| ,$$
$$g_{00}=\exp(\nu)=\left|\frac{c_1+r}{c_2 r}\right|\exp\left(\frac{-0}{2r}\right) , \quad g_{11}=-\exp(\lambda)=-\left|\frac{c_2 r}{c_1+r}\right|\exp\left(\frac{0}{2r}\right) \qquad (30)$$

The obtained solution, described by formulae (30), satisfies all equations (12), (20), (21), (22) and (25). Formulae for functions λ and ν in (30) and formulae (7), (24), (26), (28), (29), (30), describing the whole solution, do not depend on time, i.e. we obtained the stationary solution. We intentionally preserved expressions 0/r in formulae (30) because at r=0 these expressions become the indeterminate forms 0/0, but formulae (30) do not have the singularities, which appear in the standard Schwarzschild solution (16), where metric components have singularities at r=0 and $r=r_g$. Emphasize that for all r≠0, the obtained expressions for λ, ν, $g_{00}$ and $g_{11}$ in (30) coincide, to within the constants, with solution (15) for the stationary spherically symmetric gravitational field with functions $\nu(r,t)$ and $\lambda(r,t)$ without singularities.

It has already been mentioned in section 3.2, that in accordance with the Birkhoff's theorem, the spherically symmetric gravitational field in vacuum must be automatically stationary and asymptotically flat.

The obtained solution (30) means that, trying to analyze the non-stationary spherically symmetric gravitational field with functions $\nu(r,t)$ and $\lambda(r,t)$ containing singularities, we obtained A(t)=0 in (24) and (29) and respectively f(r, t)=0 in (21) and (24), i.e. the second-order successive mixed partial derivatives of functions λ(r, t) and ν(r, t) are equal to each other, see (10). It means that solution (30), describing in general case the spherically symmetric gravitational field in vacuum, is valid for both cases: for functions $\nu(r,t)$ and $\lambda(r,t)$ with and without singularities. This solution is stationary, it does not contain singularities at r=0, and coincides with the Schwarzschild solution (16) at r≠0.

Generically, the results, obtained in sections 3.5 and 3.6, once again endorse the concept that metric-affine space $G_4$ with curvature and torsion describes the distribution and motion of matter. These results also confirm logic of the general formulae (5)-(6), correctness of equations (9), (20), (22), and validity of solution (30) related to physical field with minimum curvature and torsion - the spherically symmetric stationary gravitational field in vacuum.

In order to determine functions $\nu(r,t)$, $\lambda(r,t)$ and metric components $g_{00}$, $g_{11}$ in (7), we need only formulae (30). However, in order to determine the whole geometry of metric-affine space $G_4$ with curvature and torsion, describing stationary spherically symmetric gravitational field in vacuum, we need also all $\Gamma^{i}_{jk}$ from (13) and four connections $\Gamma_{00}{}^{0}$, $\Gamma_{11}{}^{1}$, $\Gamma_{22}{}^{2}$, $\Gamma_{33}{}^{3}$ from (12).

Emphasize that using equations (5) with cosmological term, it is easy to obtain systems (14) and (20) for the spherically symmetric fields with additional "cosmological" terms. For example, the following additional terms will appear in system (20): term $-\Lambda r^2 \exp(\nu)$ in the first equation, term $-\Lambda r^2 \exp(\lambda)$ in the second equation, and term $+2r\Lambda \exp(\lambda)$ in the third equation. As result, the systems of equations with "cosmological" terms will be more complex than the respective systems (14) and (20). However, at the same time, the additional "cosmological" terms will not significantly affect the solutions, because these terms are important only over a very large scale, i.e. over the "cosmological" region of the space-time.

### 3.6. Material field in space with functions $\nu(r,t)$ and $\lambda(r,t)$ containing singularities

It has already been mentioned in section 3.6, that in the case, when there are singularities in functions $\nu(r,t)$ and $\lambda(r,t)$ (and this is the most general case), system (20) should be used to describe the spherically symmetric non-stationary field. The respective solution for the field in vacuum (the pure gravitational field) has been presented in the previous section. Now we'll try to obtain a solution for an arbitrary material spherically symmetric field in the space with functions $\nu(r,t)$ and $\lambda(r,t)$ containing singularities.

To do it, we should solve system (20), describing the material spherically symmetric field in a general case, but we cannot use now condition (12), which is valid only for the gravitational field in vacuum. In order to determine functions $\nu(r,t)$ and $\lambda(r,t)$ in system (20) and then metric components $g_{00}$ and $g_{11}$ in (7), we need only the first five equations from (20). However, to determine the whole geometry of the spherically symmetric metric-affine space $G_4$, we need all $\Gamma^{i}_{jk}$ from (8) and four unknown $\Gamma_{00}{}^{0}$, $\Gamma_{11}{}^{1}$, $\Gamma_{22}{}^{2}$, $\Gamma_{33}{}^{3}$, which should be determined from (20).

Emphasize, that using (20), the unknown $\Gamma_{00}{}^{0}$, $\Gamma_{11}{}^{1}$, $\Gamma_{22}{}^{2}$, $\Gamma_{33}{}^{3}$ cannot be determined uniquely. In other words, these four connections may have different forms, i.e. they may be related to the different spherically symmetric material fields. Recall also that, if there are no singularities in functions $\nu(r,t)$ and $\lambda(r,t)$, and therefore their second-order successive mixed derivatives are equal to each other (see (10)), then equations (20) coincide with equations (19).

Now let us try to determine the unknown functions $\nu(r,t)$ and $\lambda(r,t)$ by using equations (20). To do this, first of all, introduce a new function F(r, t):

$$F(r,t)=c\frac{\partial\,\Gamma_{11}{}^{1}(r,t)}{\partial\, t}-\frac{\partial\,\Gamma_{00}{}^{0}(r,t)}{\partial\, r}=-\frac{c}{4}\left[\frac{\partial}{\partial t}(\nu'-\lambda')+\frac{\partial}{\partial r}(\dot{\nu}-\dot{\lambda})\right] \tag{31}$$

Then substitute functions (21) and (31) in the fourth and fifth equations in (20) respectively.

Finally, we obtain the following system of equations:

$$\begin{gathered} r\lambda'-1+\exp\lambda=0\ , \\ r\nu'+1-\exp\lambda=0\ , \\ \left[\nu''+\frac{(\nu')^{2}}{2}-\frac{\nu'\lambda'}{2}+\frac{\nu'-\lambda'}{r}\right]=\frac{c^{2}}{2}\exp(\lambda-\nu)\left[\frac{f(r,t)^{2}}{2}+\frac{\partial f(r,t)}{\partial t}\right]\ , \end{gathered} \tag{32}$$

$$
\begin{gathered}
f(r,t)=\left(\dot{\lambda}-\dot{\nu}\right)=\frac{r}{2}\left[\frac{\partial}{\partial r}\left(\dot{\nu}-\dot{\lambda}\right)-\frac{\partial}{\partial t}\left(\nu'-\lambda'\right)\right] , \\
F(r,t)=c\frac{\partial\,\Gamma_{11}{}^{1}(r,t)}{\partial\,t}-\frac{\partial\,\Gamma_{00}{}^{0}(r,t)}{\partial\,r}=-\frac{c}{4}\left[\frac{\partial}{\partial t}\left(\nu'-\lambda'\right)+\frac{\partial}{\partial r}\left(\dot{\nu}-\dot{\lambda}\right)\right] , \\
c\frac{\partial\,\Gamma_{22}{}^{2}}{\partial\,t}-\frac{\partial\,\Gamma_{00}{}^{0}}{\partial\,\theta}=0 , \\
c\frac{\partial\,\Gamma_{33}{}^{3}}{\partial\,t}-\frac{\partial\,\Gamma_{00}{}^{0}}{\partial\,\varphi}=0 , \\
\frac{\partial\,\Gamma_{22}{}^{2}}{\partial\,r}-\frac{\partial\,\Gamma_{11}{}^{1}}{\partial\,\theta}=0 , \\
\frac{\partial\,\Gamma_{33}{}^{3}}{\partial\,r}-\frac{\partial\,\Gamma_{11}{}^{1}}{\partial\,\varphi}=0 , \\
\frac{\partial\,\Gamma_{33}{}^{3}}{\partial\,\theta}-\frac{\partial\,\Gamma_{22}{}^{2}}{\partial\,\varphi}=0 .
\end{gathered}
\tag{32}
$$

System (32) is simpler than system (20). Using the fourth and fifth equations in (32), it is easy to get the equation connecting functions f(r, t) and F(r, t):

$$
\frac{\partial f(r,t)}{\partial r}+\frac{1}{r}f(r,t)-\frac{2}{c}F(r,t)=0 \tag{33}
$$

Of course, it is impossible to determine two unknown functions f(r, t) and F(r, t) from one equation (33). However, there are a few physical reasoning, able to help in determining a simple relationship between these two functions:

1. In sections 3.2 and 3.3 we obtained two solutions with functions $\nu(r,t)$ and $\lambda(r,t)$ without singularities describing, respectively, the gravitational spherically symmetric field in vacuum and an arbitrary material spherically symmetric field. These two systems of equations (14) and (19) and their respective solutions were similar to each other. Therefore, it is natural to assume those two systems of equations (22) and (32) and their respective solutions with functions $\nu(r,t)$ and $\lambda(r,t)$ containing singularities but also describing the gravitational spherically symmetric field in vacuum and an arbitrary material spherically symmetric field, should be again similar to each other. Thus functions $\dfrac{f(r,t)}{r}$ and F(r, t) should be also similar to each other.

2. Analyzing equation (33) one can assume that in the simplest case functions F(r, t), $\dfrac{\partial f(r,t)}{\partial r}$ and $\dfrac{f(r,t)}{r}$ should have similar forms.

3. The fourth and fifth equations in (32) show that functions F(r, t) and $\frac{f(r,t)}{r}$ represent respectively the difference and sum of two functions $\frac{\partial}{\partial t}(\nu' - \lambda')$ and $\frac{\partial}{\partial r}(\dot{\nu} - \dot{\lambda})$. Derivatives $\frac{\partial}{\partial t}(\nu' - \lambda')$ and $\frac{\partial}{\partial r}(\dot{\nu} - \dot{\lambda})$, obtained above for the case when functions $\nu(r,t)$ and $\lambda(r,t)$ do not have singularities (see formulae (24)), show that these derivatives represent the different types of functions. Therefore, it is reasonable to assume that when functions $\nu(r,t)$ and $\lambda(r,t)$ have singularities, the derivatives $\frac{\partial}{\partial t}(\nu' - \lambda')$ and $\frac{\partial}{\partial r}(\dot{\nu} - \dot{\lambda})$ will also represent different types of functions. Thus, these derivatives will not be cancelled out during addition or subtraction, but their sum and difference will be described by similar types of functions. As result, it is again reasonable to assume that functions F(r, t) and $\frac{f(r,t)}{r}$ should be similar to each other.

Based on the above arguments, we assume that, in order to get the simplest particular solution, functions F(r, t) and $\frac{f(r,t)}{r}$ can be just proportional to each other:

$$m\frac{f(r,t)}{r} = F(r,t) \quad , \tag{34}$$

where m is the constant coefficient.

Substituting (34) in (33), it is easy to obtain the following equation for function f(r, t):

$$\frac{\partial f(r,t)}{\partial r} = \left(\frac{2m}{c} - 1\right)\frac{f(r,t)}{r} \tag{35}$$

Again taking into account that $\int \frac{dx}{x} = \ln|x|$, equation (35) can be solved and then solution of system (32) will be presented as follows:

$$|f(r,t)| = \frac{|A(t)|}{r^{(1-2m/c)}}, \quad |F(r,t)| = \frac{m|A(t)|}{r^{(2-2m/c)}}, \quad \nu = \frac{-1}{2r^{(1-2m/c)}}\int A(t)dt + \frac{1}{2}B(r) + \frac{1}{2}C(t),$$
$$\lambda = \frac{1}{2r^{(1-2m/c)}}\int A(t)dt - \frac{1}{2}B(r) + \frac{1}{2}C(t), \; \dot{\nu} = -\frac{A(t)}{2r^{(1-2m/c)}} + \frac{1}{2}\dot{C}(t),$$
$$\dot{\lambda} = \frac{A(t)}{2r^{(1-2m/c)}} + \frac{1}{2}\dot{C}(t), \; \lambda' = \frac{-(1-2m/c)}{2r^{(2-2m/c)}}\int A(t)dt + \frac{1}{2}B'(r), \tag{36}$$
$$\nu' = \frac{(1-2m/c)}{2r^{(2-2m/c)}}\int A(t)dt - \frac{1}{2}B'(r), \; \nu' - \lambda' = \frac{(1-2m/c)}{r^{(2-2m/c)}}\int A(t)dt - B'(r),$$

where A(t), B(r) and C(r) are the unknown functions.

Solution (36) satisfies equations (21), (31), (34), (35), and the first, second, fourth and fifth equations in (32). However, all functions in (36) have singularities at least at r=0.

Functions A(t), B(r) and C(r) can be determined using the third equation in (32) and the respective boundary/initial conditions. Substituting (44) in the third equation in (32), we obtain:

$$-\frac{B''(r)}{2}+\frac{[B'(r)]^2}{4}-\frac{B'(r)}{r}=-\frac{(1-2m/c)^2}{4r^{(4-4m/c)}}\left(\int A(t)dt\right)^2+$$
$$+\frac{(1-2m/c)B'(r)}{2r^{(2-2m/c)}}\int A(t)dt-\frac{(1-2m/c)m}{cr^{(3-2m/c)}}\int A(t)dt+ \qquad (37)$$
$$+\frac{c^2}{2}\left[\frac{A^2(t)}{2r^{(2-4m/c)}}+\frac{\dot{A}(t)}{r^{(1-2m/c)}}\right]\exp\left[\frac{1}{r^{(1-2m/c)}}\int A(t)dt+B(r)\right]$$

Equation (37) is similar to equation (25), and therefore it also does not contain function C(t), which is the part of general solution (36). So again, without any loss of generality, we obtain condition (26) and, respectively, the formula $\lambda=-\nu$.

Now we can use the same reasoning that has already led to the solution (30). Equation (37), containing two unknown functions A(t) and B(r), should be valid at any r and t. At the same time, the left-hand side of this equation depends only on one variable r and does not depend on t, while the right-hand side depends on both variables: r and t. This can be true only if the right-hand side of this equation does not depend on t. This is possible only if A(t)=0.

At this condition, equation (37) reduces to the general Riccati equation (27) for function B′(r). Respectively, its particular solution can be presented as (28).

Using (26), (28), (29), and (36), the final solution of system (32) and the respective metric components (7) can be presented as

$$|f(r,t)|=\frac{0}{r^{(1-2m/c)}},\ |F(r,t)|=\frac{m\cdot 0}{r^{(2-2m/c)}},\ \nu=\frac{-0}{2r^{(1-2m/c)}}-\ln\left|\left(\frac{c_2 r}{c_1+r}\right)\right|,\ \lambda=\frac{0}{2r^{(1-2m/c)}}+\ln\left|\left(\frac{c_2 r}{c_1+r}\right)\right|$$
$$g_{00}=\exp(\nu)=\left|\frac{c_1+r}{c_2 r}\right|\exp\left(\frac{-0}{2r^{(1-2m/c)}}\right),\ g_{11}=-\exp(\lambda)=-\left|\frac{c_2 r}{c_1+r}\right|\exp\left(\frac{0}{2r^{(1-2m/c)}}\right), \qquad (38)$$

where c is the speed of light, $c_1$ and $c_2$ are the new constants.

The obtained solution, described by formulae (26), (28), (29), (36) and (38), satisfies all equations (31)-(35), (37) and (27). Formulae for functions λ and ν in (38) and formulae (7), (26), (28), (29), (36) and (38), describing the whole solution, do not depend on time, i.e. we obtained the stationary solution. However now at r=0 we have expressions 0/0, which represent the indeterminate forms, but formulae (38) do not have singularities, unlike the standard Schwarzschild solution (16), where metric components have singularities at r=0 and $r=r_g$. For all $r\neq 0$, the obtained expressions for λ, ν and metric components $g_{00}$, $g_{11}$ in (38) coincide to within the constants with solution (15) for the stationary spherically symmetric gravitational field with functions λ (r, t) and ν(r, t) without singularities.

It means that obtained solution (38) describes in general case the spherically symmetric arbitrary material field. This solution is valid for both cases: for functions λ(r, t) and ν(r, t) with and without singularities.

It has already been mentioned, that in accordance with the generalized Birkhoff's theorem, the spherically symmetric material field must be automatically stationary and asymptotically flat. Thus, trying to analyze the non-stationary spherically symmetric material field with functions λ(r, t) and ν(r, t) containing singularities, we obtained that A(t)=0 in (24) and (36) and respectively f(r, t)=0 and F(r, t)=0 in (21), (31), (36), (38), i.e. the second-order successive mixed partial derivatives of functions λ(r, t) and ν(r, t) are equal to each other (see (11)).

Note again, that formulae (38), describing the particular solution based on the assumption (34), allow calculating functions $\nu(r,t)$ and $\lambda(r,t)$ in (20) and (32), and metric components $g_{00}$ and $g_{11}$ in (7). However, in order to determine the whole geometry of the metric-affine space $G_4$ with curvature and torsion, describing the spherically symmetric arbitrary material field, we need all $\Gamma^i_{jk}$ from (8) and four $\Gamma_{00}{}^0$, $\Gamma_{11}{}^1$, $\Gamma_{22}{}^2$, $\Gamma_{33}{}^3$ , which should be determined from the six last equations in (32).

Emphasize that formulae (38), (8) and (32) describe the general case – a spherically symmetric arbitrary material field, while formulae (30), (12), (13), which can be derived from (38), (8), (32) as a particular case, describe the respective particular case – the spherically symmetric gravitational field.

Note that laws of the conservation of energy and momentum in the described metric-affine space $G_4$ are identical to the respective conservation laws in the GR.

## 4. Uniform isotropic fields

### 4.1. Field equations

Now obtain the "cosmological" equations concerning evolution of the Universe as a whole, based on the assumption of the isotropy and uniformity of matter distribution in the space. It is well known that stars and galaxies are distributed over space in a very non-uniform manner, but in studying the Universe "on a large scale" one can disregard the "local" inhomogeneities produced by stars and galaxies, because in the general terms the uniform isotropic model gives an adequate description not only of the present state of the Universe, but also of a significant part of its evolution in the past [16]. At the same time it is clear that, by its very nature, the assumption of the homogeneity and isotropy of the Universe have only an approximate character, since these properties surely are not valid if we go to a smaller scale.

We'll start analysis of equations (5)-(6) for the uniform isotropic space using at first the closed uniform isotropic model of the Universe. For this model, the non-zero components of metric tensor in the 4D "spherical" coordinates $(\eta,\chi,\theta,\varphi)$ equal [16]

$$g_{00}=a^2,\ g_{11}=-a^2,\ g_{22}=-a^2\sin^2\chi,\ g_{33}=-a^2\sin^2\chi\sin^2\theta,\ g_{ik}g^{kl}=\delta_i^l, \tag{39}$$

where $a(\eta)$ is the radius of the space curvature depending on the coordinate $\eta$ used for convenience in place of the time coordinate.

Radius $a(\eta)$ is actually a radius of a hyper-sphere, it cannot be negative, and quantity $\eta$ is related to time t as follows [16]:

$$c\,dt=a(\eta)d\eta, \tag{40}$$

where $c$ is the speed of light.

Due to the uniformity and isotropy of the space, the number of equations in system (5)-(6) and number of unknowns are much less than the respective numbers in a general case.

Substituting (39) in (6), we obtain the following connections $\Gamma_{jk}^{\ \ i}$ :

$$\Gamma_{01}^{\ \ 0}=\Gamma_{11}^{\ \ 1},\ \Gamma_{11}^{\ \ 0}=\Gamma_{01}^{\ \ 1}=\Gamma_{02}^{\ \ 2}=\Gamma_{03}^{\ \ 3}=\frac{a'}{a},\ \Gamma_{22}^{\ \ 0}=\frac{a'}{a}\sin^2\chi\ ,\ \Gamma_{33}^{\ \ 0}=\frac{a'}{a}\sin^2\chi\sin^2\theta\ ,$$

$$\Gamma_{02}^{\ \ 0}=\Gamma_{12}^{\ \ 1}=\Gamma_{22}^{\ \ 2},\ \Gamma_{03}^{\ \ 0}=\Gamma_{13}^{\ \ 1}=\Gamma_{23}^{\ \ 2}=\Gamma_{33}^{\ \ 3},\ \Gamma_{10}^{\ \ 1}=\Gamma_{20}^{\ \ 2}=\Gamma_{30}^{\ \ 3}=\Gamma_{00}^{\ \ 0}\ ,$$

$$\Gamma_{22}^{\ \ 1}=-\sin\chi\cos\chi\ ,\ \Gamma_{33}^{\ \ 1}=-\sin\chi\cos\chi\sin^2\theta\ ,\ \ \Gamma_{12}^{\ \ 2}=\Gamma_{13}^{\ \ 3}=\cot\chi\ , \tag{41}$$

$$\Gamma_{21}^{\ \ 2}=\Gamma_{31}^{\ \ 3}=\cot\chi+\Gamma_{11}^{\ \ 1},\ \Gamma_{33}^{\ \ 2}=-\sin\theta\cos\theta,\ \Gamma_{23}^{\ \ 3}=\cot\theta,\ \Gamma_{32}^{\ \ 3}=\cot\theta+\Gamma_{22}^{\ \ 2},$$

where four connections $\Gamma_{00}^{\ \ 0}$, $\Gamma_{11}^{\ \ 1}$, $\Gamma_{22}^{\ \ 2}$, $\Gamma_{33}^{\ \ 3}$ are undetermined, all other $\Gamma_{jk}^{\ \ i}$ are equal to zero, and the prime on function $a(\eta)$ denotes differentiation with respect to $\eta$.

Four undetermined connections $\Gamma_{00}^{\ \ 0}$, $\Gamma_{11}^{\ \ 1}$, $\Gamma_{22}^{\ \ 2}$, $\Gamma_{33}^{\ \ 3}$ appear in solution (41) because the number of the independent equations in system (5)-(6) in the uniform isotropic field, unlike any general non-uniform and non-isotropic case, is less than the number of the variables. Note, that this indefiniteness is probably similar to the one described in [18].

Using (5), (39) and (41) we obtain the following system of equations describing the closed model of the uniform isotropic field in $G_4$ space

$$R_{00}-\frac{1}{2}g_{00}R=3\left(\frac{a'}{a}\right)^2+3-\Lambda a^2=0\ ,$$

$$R_{11}-\frac{1}{2}g_{11}R=\left(\frac{a'}{a}\right)^2-2\frac{a''}{a}-1-\Lambda a^2=0\ ,$$

$$R_{22}-\frac{1}{2}g_{22}R=\left[\left(\frac{a'}{a}\right)^2-2\frac{a''}{a}-1-\Lambda a^2\right]\sin^2\chi=0\ , \tag{42}$$

$$R_{33}-\frac{1}{2}g_{33}R=\left[\left(\frac{a'}{a}\right)^2-2\frac{a''}{a}-1-\Lambda a^2\right]\sin^2\chi\sin^2\theta=0\ ,$$

$$R_{01}-\frac{1}{2}g_{01}R\equiv R_{01}=-R_{10}+\frac{1}{2}g_{10}R\equiv -R_{10}=\frac{\partial\Gamma_{11}^{\ \ 1}}{\partial\eta}-\frac{\partial\Gamma_{00}^{\ \ 0}}{\partial\chi}=0\ ,$$

$$
\begin{aligned}
R_{02}-\frac{1}{2}g_{02}R \equiv R_{02} &= -R_{20}+\frac{1}{2}g_{20}R \equiv -R_{20} = \frac{\partial \Gamma_{22}{}^{2}}{\partial \eta}-\frac{\partial \Gamma_{00}{}^{0}}{\partial \theta}=0 \quad , \\
R_{03}-\frac{1}{2}g_{03}R \equiv R_{03} &= -R_{30}+\frac{1}{2}g_{30}R \equiv -R_{30} = \frac{\partial \Gamma_{33}{}^{3}}{\partial \eta}-\frac{\partial \Gamma_{00}{}^{0}}{\partial \varphi}=0 , \\
R_{12}-\frac{1}{2}g_{12}R \equiv R_{12} &= -R_{21}+\frac{1}{2}g_{21}R \equiv -R_{21} = \frac{\partial \Gamma_{22}{}^{2}}{\partial \chi}-\frac{\partial \Gamma_{11}{}^{1}}{\partial \theta}=0 \quad , \\
R_{13}-\frac{1}{2}g_{13}R \equiv R_{13} &= -R_{31}+\frac{1}{2}g_{31}R \equiv -R_{31} = \frac{\partial \Gamma_{11}{}^{1}}{\partial \varphi}-\frac{\partial \Gamma_{33}{}^{3}}{\partial \chi}=0 \quad , \\
R_{23}-\frac{1}{2}g_{23}R \equiv R_{23} &= -R_{32}+\frac{1}{2}g_{32}R \equiv -R_{32} = \frac{\partial \Gamma_{22}{}^{2}}{\partial \varphi}-\frac{\partial \Gamma_{33}{}^{3}}{\partial \theta}=0 \quad ,
\end{aligned}
\tag{42}
$$

where the double prime on function $a(\eta)$ denotes double differentiation with respect to $\eta$.

Note that the third and fourth equations in (42) are similar to the second one; it means they are not independent, and therefore can be disregarded. In addition emphasize, that the first two equations in (42) can be easily reduced to the more simple equations by substituting $a'/a$ from the first equation in the second equation. As result, instead of (42), we obtain the following rather simple system of equations:

$$
\begin{aligned}
&\left(\frac{a'}{a}\right)^{2}+1-\frac{\Lambda}{3}a^{2}=0 \\
&\frac{a''}{a}+1-\frac{2\Lambda}{3}a^{2}=0 , \\
&\frac{\partial \Gamma_{11}{}^{1}}{\partial \eta}-\frac{\partial \Gamma_{00}{}^{0}}{\partial \chi}=0 , \\
&\frac{\partial \Gamma_{22}{}^{2}}{\partial \eta}-\frac{\partial \Gamma_{00}{}^{0}}{\partial \theta}=0 , \\
&\frac{\partial \Gamma_{33}{}^{3}}{\partial \eta}-\frac{\partial \Gamma_{00}{}^{0}}{\partial \varphi}=0 , \\
&\frac{\partial \Gamma_{22}{}^{2}}{\partial \chi}-\frac{\partial \Gamma_{11}{}^{1}}{\partial \theta}=0 , \\
&\frac{\partial \Gamma_{11}{}^{1}}{\partial \varphi}-\frac{\partial \Gamma_{33}{}^{3}}{\partial \chi}=0 , \\
&\frac{\partial \Gamma_{22}{}^{2}}{\partial \varphi}-\frac{\partial \Gamma_{33}{}^{3}}{\partial \theta}=0 \ ,
\end{aligned}
\tag{43}
$$

where $a'/a=H(\eta)$ is actually the Hubble parameter measuring the expansion rate.

System (43) shows that in order to obtain radius $a(\eta)$ of the space curvature, we should solve only the first two equations. But to determine the whole geometry of the uniform isotropic space G4, we have to obtain four connections $\Gamma_{00}{}^{0}$, $\Gamma_{11}{}^{1}$, $\Gamma_{22}{}^{2}$, $\Gamma_{33}{}^{3}$ from the last six equations in (43) and twelve $\Gamma_{jk}{}^{i}$ depending on $\Gamma_{00}{}^{0}$, $\Gamma_{11}{}^{1}$, $\Gamma_{22}{}^{2}$, $\Gamma_{33}{}^{3}$ from (41), all other $\Gamma_{jk}{}^{i}$ equal to zero.

Note that since each of $\Gamma_{00}{}^{0}$, $\Gamma_{11}{}^{1}$, $\Gamma_{22}{}^{2}$, $\Gamma_{33}{}^{3}$ depends on four variables $(\eta,\chi,\theta,\varphi)$, the last six equations in (43) do not allow determining uniquely these four connections. This means, that connections $\Gamma_{00}{}^{0}$, $\Gamma_{11}{}^{1}$, $\Gamma_{22}{}^{2}$, $\Gamma_{33}{}^{3}$ may have various forms related to the different uniform isotropic material fields.

Emphasize that first two equations in (43), determining the space curvature radius $a(\eta)$, are valid only if $a(\eta)\neq 0$, because when $a(\eta)=0$ the first terms in these equations become indeterminate or even infinite, i.e. these equations have the singular points. However, at the condition that $a(\eta)\neq 0$, the first two equations in (43) can be presented as

$$(a')^2 + a^2 - \frac{\Lambda}{3}a^4 = 0 \ ,$$
$$a'' + a - \frac{2\Lambda}{3}a^3 = 0 . \qquad (44)$$

Recall that equations (44) are valid only for the closed uniform isotropic model of Universe.

To obtain the respective equations, describing the open uniform isotropic model, we should, according to [16], replace $\eta,\chi,a$ in all formulae (39)-(44) respectively with $i\eta,\, i\chi,\, ia$. Now the radius $ia(\eta)$ of the space curvature becomes an imaginary radius of a pseudo-sphere. As result, instead of (44), we obtain the following system of two equations for the open model:

$$(a')^2 - a^2 - \frac{\Lambda}{3}a^4 = 0$$
$$a'' - a - \frac{2\Lambda}{3}a^3 = 0 . \qquad (45)$$

Similar to equations (44), this system is also valid only if $a(\eta)\neq 0$, because the respective equations in the general system, describing the open uniform isotropic model of the Universe, has (similar to the first two equations in (43) for the closed model) the singularities at $a(\eta)=0$.

**4.2. Gravitational field in vacuum**

To solve system (51) in the simplest case, we should make an assumption regarding four undetermined connections $\Gamma_{00}{}^{0}$, $\Gamma_{11}{}^{1}$, $\Gamma_{22}{}^{2}$, $\Gamma_{33}{}^{3}$. The simplest solution can be again obtained by using condition (12). Its substitution in (41) yields

$$\Gamma_{11}{}^{0}=\Gamma_{01}{}^{1}=\Gamma_{02}{}^{2}=\Gamma_{03}{}^{3}=\frac{a'}{a},\quad \Gamma_{22}{}^{0}=\frac{a'}{a}\sin^{2}\chi,\ \Gamma_{33}{}^{0}=\frac{a'}{a}\sin^{2}\chi\sin^{2}\theta,$$

$$\Gamma_{22}{}^{1}=-\sin\chi\cos\chi,\ \Gamma_{33}{}^{1}=-\sin\chi\cos\chi\sin^{2}\theta,\ \Gamma_{12}{}^{2}=\Gamma_{21}{}^{2}=\Gamma_{13}{}^{3}=\Gamma_{31}{}^{3}=\cot\chi, \qquad (46)$$

$$\Gamma_{33}{}^{2}=-\sin\theta\cos\theta,\ \Gamma_{23}{}^{3}=\Gamma_{32}{}^{3}=\cot\theta,\ \text{all other } \Gamma_{jk}{}^{i} \text{ are equal to zero}$$

Solution (46) is related to the minimal curvature and torsion of the uniform isotropic space-time, i.e. to the uniform isotropic gravitational field in vacuum. Recall, that different sets of $\Gamma_{jk}{}^{i}$ satisfying (41), are related to various degrees of the space-time curvature and torsion, and describe the different material fields. The simplest set (46) of connections $\Gamma_{jk}{}^{i}$ obtained by using condition (12), describes the "simplest" physical field, i.e. the uniform isotropic gravitational field in vacuum away from the gravitating bodies.

It is obvious that using formulae (12), system (43) for the closed model can be reduced to two equations (44) for the radius $a(\eta)$ of the space curvature. Note that second equation in (44), describing the closed isotropic model, is similar to the respective equation for the dust-like matter in the GR [16].

The first equation in (44) is the first order autonomous equation, and its general solution $a(\eta)$ can be obtained as a solution of the integral equation

$$\eta=\int\pm\left[\frac{\Lambda}{3}a^{4}-a^{2}\right]^{-1/2}da+C_{1}\ , \qquad (47)$$

where $C_1$ is an arbitrary constant.

The second equation in (44) is the second-order autonomous equation, and its general solution $a(\eta)$ can be obtained as the solution of the integral equation

$$\pm\eta=\int\left[C_{2}+2\int\left(\frac{2\Lambda}{3}a^{3}-a\right)da\right]^{-1/2}da-C_{3}, \qquad (48)$$

where $C_2$ and $C_3$ are the constants.

Integrals in equations (47)-(48) can be calculated via the hyper-geometric functions. As result, the final formula for solution $a(\eta)$ of equations (44) has a complex form, which is difficult to analyze. This conclusion is also applicable to equations (53). However, our solution should satisfy simultaneously both equations in (44) or both equations in (45). Therefore, our final solution can be only the particular one satisfying simultaneously both equations in systems (52) or (53). It will be discussed below in section 4.4. Let us start our analysis from the simplified systems of equations (44) and (45) and obtain their approximate analytical solutions.

### 4.3. Solution without cosmological term for gravitational field in vacuum

Since the value of cosmological constant $\Lambda\approx10^{-52}\text{m}^{-2}$ is very small, we can neglect terms $-\frac{1}{3}\Lambda a^{4}$ and $-\frac{2}{3}\Lambda a^{3}$ in the equations (44)-(45) when $a(\eta)$ is not too large, i.e. the space-time is not too curved and torsional, and the gravitational field is not too strong. As result, we obtain two following simplified systems of equations:

For the closed model:

$$\begin{aligned} (a')^2 + a^2 &= 0 \\ a'' + a &= 0\,, \end{aligned} \qquad (49)$$

For the open model:

$$\begin{aligned} (a')^2 - a^2 &= 0 \\ a'' - a &= 0 \end{aligned} \qquad (50)$$

The first equations in both systems (49) and (50) are the simple autonomous equations, while the second equations (also in both systems) are the equations of free oscillations. The general solutions of these types of equations are well-known. The problem is to find a solution, satisfying both equations simultaneously.

System (49) for the closed model has the following general solution satisfying both equations in (49):

$$a(\eta) = \left| A_1 \exp[(B_1 + iC_1)\eta] \right|, \qquad (51)$$

where $A_1$, $B_1$ and $C_1$ are the constants.

Modulus in the right hand side in (51) and other formulae below in this section appear because radius $a(\eta)$ of the space curvature can be only real and positive. (Recall, that probably only modulus of a complex function is physically meaningful, because it is measurable).

Coefficients $B_1$ and $C_1$ in (51) are connected by equations

$$(B_1 + iC_1)^2 = -1 \qquad \text{i.e.} \qquad (B_1 + iC_1) = \pm i\,, \qquad (52)$$

which can be easily obtained substituting (51) in (49).

Equations (52) mean that expression $(B_1 + iC_1)$ is the imaginary one. Using it and formulae (51) and (40), we can space curvature radius $a(t)$ for closed model:

$$\begin{gathered} a(\eta) = \left| A_1 \exp[(B_1 + iC_1)\eta] \right| = A_1 \left| \exp(\pm iC_1\eta) \right| \\ ct = \int \left| A_1 \exp[(B_1 + iC_1)\eta] \right| d\eta = A_1 \int \left| \exp(\pm iC_1\eta) \right| d\eta = A_1 \left| \frac{\exp(\pm iC_1\eta)}{\pm i} \right| + M\,, \\ a(t) = M \left| \pm ict) \right| = Mc|t| + N \end{gathered} \qquad (53)$$

where M and N are the constants, and c is the speed of light.

Formulae (53), describing the approximate solution without cosmological term for the gravitational field in vacuum, show that in the closed model the Universe undergoes a single transition from the contraction to expansion, i.e. it performs the non-singular classical bounce, where radius $a(t)$ of the space curvature linearly decreases during the contraction phase and then linearly increases during the expansion phase. Note also that obtained solution (53) does not have problem of the initial conditions because the contraction phase of the Universe development started in the infinite past.

For the open model we obtain the following general solution satisfying both equations (50):

$$a(\eta)=\left|A_2 \exp[(B_2+iC_2)\eta]\right| \quad , \qquad (54)$$

where $A_2$, $B_2$ and $C_2$ are the constants.

Coefficients $B_2$ and $C_2$ in (54) are connected by equations

$$(B_2+iC_2)^2=+1, \qquad \text{i.e.} \qquad (B_1+iC_1)=\pm 1 \quad , \qquad (55)$$

which can be easily obtained substituting (54) into (50).

Equations (55) mean that expression $(B_2+iC_2)$ is the real one. Using this result and formulae (54) and (40), we can determine radius $a(t)$ of the space curvature for the open model:

$$a(\eta)=\left|A_2 \exp[(B_2+iC_2)\eta]\right| = A_2 \exp(\pm B_2\eta) \ ,$$

$$ct=\int\left|A_2 \exp[(B_2+iC_2)\eta]\right|d\eta = A_2\int \exp(\pm B_2\eta)d\eta = \pm\, A_2 \exp(\pm B_2\eta)+N \ , \qquad (56)$$

$$a(t)=Mc|t|+N$$

where M and N are the constants, and c is the speed of light.

Formulae (56), describing the approximate solution without cosmological term for the gravitational field in vacuum, show that in the open model, similar to formulae (53) for the closed model, the Universe undergoes a single transition from the contraction to expansion, i.e. it performs the non-singular classical bounce, where radius $a(t)$ of the space curvature linearly decreases during the contraction phase and then linearly increases during the expansion phase. Note also that obtained solution (56) does not have problem of the initial conditions because the contraction phase of the Universe development started in the infinite past.

It is necessary to emphasize again that, as it follows from formulae (53) and (56), in both solutions (for the closed and open models) the curvature radius $a(t)$ cannot be equal to zero. In other words, solutions (53) and (56) do not have singularities.

**4.4. Solutions of equations with cosmological term for gravitational field in vacuum**

Now solve equations (44) and (45) with cosmological terms describing gravitational fields in vacuum (i.e. fields with the minimal curvature and torsion) in the closed and open uniform isotropic models of Universe.

Start from the general solution (47) of the first equations in (44) and (45), which are the first order autonomous equations. The general solution $a(\eta)$ of integral equation (47) for the closed model can be obtained in comparatively simple form as follows

$$a(\eta)=\left|\frac{\pm 2C_1\sqrt{3}\exp(\pm i\eta)}{\Lambda C_1^{\ 2}-\exp(\pm 2i\eta)}\right| \quad , \tag{57}$$

where $C_1$ is the constant.

Note that solving the first order autonomous equation, we used constant in the form $\ln(C_1\sqrt{\Lambda})$, because it significantly simplifies further calculations.

The general solution $a(\eta)$ of integral equation (47) for the open model can be obtained in a similar way as the solution of the first equation in (45):

$$a(\eta)=\left|\frac{\pm 2C_3\sqrt{3}\exp(\pm\eta)}{\Lambda C_3^{\ 2}-\exp(\pm 2\eta)}\right| \quad , \tag{58}$$

where $C_3$ is the constant.

The correct signs in (57) and (58) and all other formulae below in this section should be selected based on the physical reasoning.

In the limiting cases when cosmological constant $\Lambda=0$, expressions (57) and (58) are reduced, respectively, to formulae (53) and (56) for $a(\eta)$.

Recall that the general solution of the second equations in (44)-(45) have complex forms (see (48)) and can be expressed via the elliptic integrals of the first kind and inverse hyperbolic sin functions. However, the obtained general solutions (57) and (58) of the first equations in systems (44) and (45) also satisfy the second equations in these systems (it can be proved by substitution). Thus, formulae (57) and (58) present also the particular solutions of the second order autonomous equations in (44) and (45), i.e. solutions of the integral equation (48). So generically, solutions (57) and (58) are the particular solutions of systems (44) and (45) respectively.

However, each of the solutions, (57) and (58), has only one arbitrary constant, while the second order differential equation (i.e. each of the second equations in systems (44) and (45)) should contain two arbitrary constants in the general solution. Therefore, the particular solutions (57)-(58), satisfying simultaneously both equations in systems (44) and (45), should have two arbitrary constants. The second constant can be introduced as an additional coefficient in the amplitude of the particular solutions (57)-(58). This is a typical presentation of the arbitrary constants used during the solutions of differential equations.

Therefore, we obtain instead of (57) the following solution containing two constants $C_1$ and $C_2$:

$$a(\eta)=\left|\pm\frac{1}{C_2}\frac{2C_1\sqrt{3}\exp(\pm i\eta)}{\Lambda C_1^{\ 2}-\exp(\pm 2i\eta)}\right| \tag{59}$$

Selection of the second constant in the form $1/C_2$ in (59) will be convenient for the future calculations.

Note that for the closed model, where we are using the complex functions, the constant $C_2$ should be also chosen as a complex number.

For the open model we have respectively, instead of (50), the following solution containing two constants $C_3$ and $C_4$:

$$a(\eta)=\left|\pm\frac{1}{C_4}\frac{2C_3\sqrt{3}\exp(\pm\eta)}{\Lambda C_3^{\ 2}-\exp(\pm 2\eta)}\right| \tag{60}$$

Expressions (59) and (60) in the limiting case when $\Lambda \to 0$ are reduced respectively to formulae (53) and (58) to within the constants.

Using formulae (59) and (40), we can determine radius $a(t)$ of the space curvature for the gravitational field in vacuum in the closed model:

$$\sqrt{\Lambda/3}\,ct=\ln\left(\left|\frac{\exp(\pm i\eta)+C_1\sqrt{\Lambda}}{\exp(\pm i\eta)-C_1\sqrt{\Lambda}}\right|\right),\qquad a(t)=\left|\frac{\pm\sqrt{3}}{2C_1\sqrt{\Lambda}}\frac{1-\exp(\pm 2i\sqrt{\Lambda/3}cC_2t)}{\exp(\pm i\sqrt{\Lambda/3}cC_2t)}\right|\ , \tag{61}$$

where $C_1$ and $C_2$ are the constants and c is the speed of light.

Using formulae (60) and (40), radius $a(t)$ of the space curvature for the gravitational field in vacuum can be determined in the open model:

$$\sqrt{\Lambda/3}\,ct=\ln\left(\left|\frac{\exp(\pm\eta)+C_3\sqrt{\Lambda}}{\exp(\pm\eta)-C_3\sqrt{\Lambda}}\right|\right),\qquad a(t)=\left|\frac{\pm\sqrt{3}}{2C_3\sqrt{\Lambda}}\frac{1-\exp(\pm 2\sqrt{\Lambda/3}cC_4t)}{\exp(\pm\sqrt{\Lambda/3}cC_4t)}\right|, \tag{62}$$

where $C_3$ and $C_4$ are the constants and c is the speed of light.

The first expressions, connecting t and $\eta$ in formulae (61) and (62), become the identities 0=0 in the limiting case when $\Lambda=0$.

The second expressions for $a(t)$ in (61) and (62) in the limiting case when cosmological constant $\Lambda \to 0$, give the indeterminate forms 0/0. Applying the L'Hopital's rule and differentiating numerators and denominators of the second expressions for $a(t)$ in (61) and (62) with respect to $\Lambda$, we can convert them into the expressions $a(t)=Ac|t|+B$, which coincide with formulae (53) and (56) to within the constants.

Analyze formulae (61) and (62) starting from formula (62) for the open model, because it contains only the real functions, and therefore analysis of (62) is rather simple. Formula (62) shows that in the open model of Universe the radius $a(t) \to +\infty$ in two cases, when $t \to -\infty$, i.e. at the initial phase of the Universe contraction, and when $t \to +\infty$, i.e. at the final phase of the Universe expansion. In these extreme cases we have the flat, non-curved and non-torsional Euclidean space. Moreover, the metric (39) in these cases can be reduced to the Galilean metric (17). Between these two mentioned extreme cases, the radius of curvature $a(t)$ has finite values and therefore $a(t)$=min at the some critical value $t=t_{crit}$. This is a moment of the Big Bang/Big Crunch (or the Big Bounce moment). In other words, this open space-time model performs a single classical transition (bounce) from the contraction to expansion, where radius $a(t)$ of the space curvature exponentially decreases during the contraction phase, then exponentially increases during the expansion phase, and does not have a singularity.

Now analyze solution (61) for the closed model. This formula shows that in the closed model radius $a(t)$ oscillates, repeating periods of the expansion and contraction, and has only finite values without singularities. Modulus (61) of curvature radius $a(t)$ oscillates with variable amplitudes in different cycles, because function $\left|\exp\left(\pm i\sqrt{\Lambda/3}cC_2t\right)\right|$, where $C_2$ is the complex number, varies in the phase and amplitude according to the approximate formula derived from (61)

$$|a(t)| = \frac{1}{2C_1}\sqrt{\frac{3}{\Lambda}}\exp(-c|C_2|\sqrt{\frac{\Lambda}{3}}|t|)\sqrt{2-2\cos(2c|C_2|\sqrt{\frac{\Lambda}{3}}|t|)} \tag{63}$$

This result, to some extent, is similar to the results obtained using hypothesis of the cyclic Universe [21].

Note that formulae (53), (57), (59) and (61), describing the complex functions $a(\eta)$ and $a(t)$ of the space-time curvature radius in the closed model, can be also used to get the real parts of both complex functions for the additional characterization of this space-time in the closed model. Obtained results are similar to the respective formulae, presented above in this section and describing moduli of complex functions $a(\eta)$ and $a(t)$.

**4.5. Discussion of the obtained results**

Combining the results, obtained for two cases (the general solutions (53) and (56) of the approximate equations (49)-(50) without cosmological terms and the particular solutions (61)-(62) of the exact equations (44)-(45) with cosmological terms), describing the gravitational fields in vacuum, which have the minimal curvature and torsion, we can conclude the following.

The radius $a(t)$ in the closed model oscillates with variable amplitudes in different cycles, repeating periods of the expansion and contraction, and has only the finite values without singularities, see (61). This solution reduces in the limiting case $\Lambda$=0 to the respective approximate solution (53).

Formula (61) shows that in the closed model the Universe develops through three basic phases (expansion, contraction, and bounce) that are repeated from cycle to cycle with various amplitudes, i.e. radius $a(t)$ oscillates. Thus, the evolution of the cyclic Universe is dominantly classical; it is infinitely old and endures forever. In other words, it has multiple non-singular classical Big Bounces, i.e. performs periodical transitions from the contraction phase to expansion phase, where $a(\eta)$ has no singularities. The moments, corresponding to the ends/beginnings of the periodical cycles, are the initial moments of the periodical stages, when the Universe begins the expansion, and/or the final moments of the periodical stages, when the Universe ends the contraction. Such periodical moments can be called the Big Bang/Big Crunch moments (or the Big Bounce moments). These moments are not the singular points of the space-time metric $g_{ik}$ of the closed uniform isotropic model because the curvature radius $a$ in (44)-(45) cannot be equal to zero. The described cyclic processes of the Universe development are infinitely old and are repeated with time permanently. It occurs because solution (61) describes only the uniform and isotropic pure gravitational field in vacuum; such field of course is not usually admissible in the real situations, but this solution has a considerable methodological interest.

Emphasize again that obtained results for the closed model, to some extent, are similar to the results based on the hypothesis of the cyclic (oscillatory) ekpyrotic Universe, see e.g. [21] and references therein. Authors of some cyclic theories, e.g. [21], claim that there is more expansion than contraction in each cycle; and that is why the space grows exponentially from cycle to cycle. This implies that successive cycles grow longer and larger; it allows incorporating the general consequences of an increase of entropy from cycle to cycle, in accordance with the second law of thermodynamics. Note, that it would be more appropriate to call such models not cyclic but “spiral”.

The cyclic models have a few big advantages over the inflationary models: they show that smoothing and flattering occur during the contraction phase rather than the inflation phase, explain why there must be the dark energy, predict that current accelerated expansion must eventually stop and vacuum must decay in order to initiate the next cycle, and so on.

However, at the same time, it is reasonable to assume that cycles in the spiral model cannot continue indefinitely into the past, because they would become smaller than the smallest finite-sized elementary particles [22]. Moreover, as the cycles continue to increase in size with time into the future, the oscillating Universe appears increasingly ‘flat’, although it should be closed with positive spatial curvature [22]. Extrapolating back in time, cycles before the present one, become shorter and smaller culminating in another Big Bang phenomenon, and thus not replacing it. It will also bring back many old problems related to the “classical” Big Bang, such as a possible singularity during a Big Bang, inability to describe the Universe before Big Bang, and so on. Probably, the dark energy can help in this puzzling situation by providing a new hope for a consistent “cyclic-spiral” cosmology. There are many different cyclic models of the Universe, e.g. the ekpyrotic model, brane cosmology model, and so on.

The other possibility to solve this problem is to take into account that, when the large regions of the Universe are considered, the gravitational fields become important [23], because they change the space-time metric. The metric properties of the space-time may in a sense be regarded as the "external conditions", to which the bodies are subject. Thermodynamic statement that a closed system must, over a sufficiently long time, reach a state of equilibrium, applies only to the system in the steady external conditions. However, when the space-time metric varies with time, the "external conditions" of course are not steady [23]. At the same time, the gravitational field cannot itself be included in the closed system, since the conservation laws would then be reduced to the identities. For this reason, in the GR, the Universe as a whole must be regarded not as a closed system but as a system in a variable gravitational field. Therefore, application of the law of increase of entropy does not prove that the statistical equilibrium must necessarily exist [23].

Now analyze the results obtained for the open model. In this model radius $a(t)$ of the space curvature exponentially decreases during the contraction phase and then exponentially increases during the expansion phase, and does not have a singularity, see formula (62). This solution reduces in the limiting case $\Lambda=0$ to the respective approximate solution (56).

Since in both models (the closed and the open ones) the Universe is infinitely old, the obtained solutions (61)-(62) do not have problems of the initial conditions, because the cycles in the closed model and the contraction phase in the open model continue to the infinite past.

Generically, the results, obtained above in sections 4.1-4.4, endorse the concept that metric-affine space $G_4$ with curvature and torsion describes the distribution and motion of matter. These results also confirm logic of the general formulae (5)-(6), validity of equations (44)-(45), describing cases with cosmological terms, approximate equations (49)-(50), describing cases without cosmological terms, and correctness of their respective solutions (53), (56) and (61)-(62) related to the physical fields with minimal curvature and torsion, i.e. to the uniform isotropic gravitational fields in vacuum and even to the material fields, because these solutions do not contain singularities, and are similar (at least, qualitatively) to the respective solutions, describing other models of the Universe: the cyclic closed model, Friedmann model, de Sitter model, Big Bounce open model, and "dust-like" matter in the closed model.

For example, we can use formula (59) for modulus $|a(\eta)|$ for the closed model together with formula (40). As result, we'll obtained function $|a(t)|$ in the parametric form. It will be qualitatively similar to the Friedmann's solution for the space with matter in the closed model and/or to the solution, describing the "dust-like" matter in the closed model [16].

Emphasize, that obtaining results, presented in sections 4.1-4.4, we did not invoke the extra dimensions, quantum-to-classical transition, random quantum fluctuations, branes, and other elements inspired by the string theory.

Note again, that all obtained formulae (44)-(62) determine curvature radius $a(\eta)$ of the uniform isotropic gravitational field in vacuum, while in order to determine the whole geometry of this field we should also use formulae (39), (12) and (46). Recall also, that all formulae in sections 4.2-4.4 are valid only for the uniform isotropic gravitational field in vacuum, away from the bodies and other physical fields. Therefore, these formulae cannot be used to describe an ordinary matter presenting various elementary particles and fields. Of course, the influence of gravitational field in any Universe model is very important, but this field is just a change of the space-time metric, and it does not present all properties of the Universe. Therefore, to describe the Universe containing various material fields, we should use general formulae (5)-(6).

Our formulae (53), (56), (61) and (62) show that all these solutions contain the contraction phase (or even many contraction phases), where the entropy and statistical equilibrium decrease, which at the first site contradicts the second law of thermodynamics. However, we have already mentioned that all these formulae are approximate: they describe only the uniform and isotropic gravitational field in vacuum without any material bodies and fields, while in order to properly describe the Universe behavior, we should take into account all different material bodies and fields. Moreover, the different types of fields can be converted into each other, including of course the gravitational energy, which can be converted into the matter and various types of radiation. Therefore, during the contraction phases in our solutions the entropy decreases, because we did not take into account the material bodies and fields. However, even if we could include them, one should take into account the unsteady "external conditions" of the Universe, already mentioned above, i.e. the gravitational field does not allow considering the Universe as a closed system [23]. Therefore, the increase of the entropy and statistical equilibrium of such a system must not be necessary.

**4.6. Material fields**

To determine the correct type of the material cosmological model, we have to analyze the uniform isotropic distribution of the material field. To do this, one must solve equations (43) and (41) without condition (12), since this condition is valid only for the gravitational field in vacuum.

At the same time, the results, obtained above, show that radius $a(t)$ of the space curvature can be determined using only the first two equations in (43) or equations (44)-(45). It means that the curvature radius of the Universe depends only on the gravitational field and does not depend on the matter and its "local" concentrations like planets, stars and galaxies. It also means that curvature radius $a(t)$ of the space filled with uniform isotropic matter, is determined by equations (44)-(45), (49)-(50) and their respective solutions (50)-(62), presented above for the gravitational field in vacuum. To determine the whole geometry of the arbitrary material uniform isotropic field, we should use formulae (39), (43) and (41).

A general approach based on the curved and torsional space-time, can probably help to explain some cosmological problems, when the Universe with torsion undergoes a Big Crunch/Big Bang, because during a Big Crunch phase the Universe collapses not to the point of singularity, but to a point before that; and such Big Bounce model does not have a singularity [21, 24]. One of possible explanations is a coupling between the torsion and Dirac spinors, which generates a spin-spin interaction that is significant in fermionic matter at extremely high densities. Such interaction averts the unphysical Big Bang singularity and replaces it with a cusp-like bounce before which the Universe was contracting. The rapid expansion immediately after the Big Bounce explains why the present Universe at largest scales appears spatially flat, homogeneous and isotropic.

There are many other cosmological hypotheses, where the space-time with torsion allows clarifying different issues. For example, according to one of them, the generation and evolution of the quantum fluctuations (i.e. particle creation) can be induced by the torsion from the inflation phase of the Universe development to the present days [25]. The other theory [26] explains the acceleration of expansion of the Universe without unknown dark energy and also the formation of galaxy structure without the dark matter by using just the non-symmetric gravitational theory with torsion.

One more example is based on the assumption that our observable Universe, in turn, can be treated as a "closed world", i.e. some "closed" region of the infinite and eternal Multiverse, which is the most general formation in the nature. It is possible, that our observable Universe is not an exception, and there are many other Universes in the "inflating" Multiverse, see [27].

It is obvious that the Universe structure and development are determined by parameters of the elementary particles and fundamental physical constants. In other words, even existence of the observer, according to the Anthropic principle, is determined by the current values of physical constants. Probably, the Universe itself has selected among the numerous possibilities, the values of various constants, which provide its creation and existence in the current form. One can assume that there is some similarity with concept of the "biological natural selection".

Recall that Anthropic principle is the proposition that all observations about the Universe are possible only if Universe is capable of developing observers, i.e. if the laws of nature, fundamental constants, and parameters of the Universe have values that are consistent with conditions for life. In other words, the Universe should be finely tuned to be compatible with life and the observer existence. It means that fundamental constants and conditions, necessary for life, are incredibly precise; see e.g. [28] and references therein.

All these various and highly speculative remarks address a number of the well-known cosmological issues, and have the wide implications for the fundamental physics. Hopefully, the calculations performed above, and the respective physical reasoning can help to clarify some issues related to the cosmology.

In sections 4.3-4.4 we solved equations (43) using condition (12), for the simplest uniform isotropic case and obtained solution for the uniform isotropic gravitational field in vacuum. Now we will solve these equations for the simple material fields. All non-zero connections $\Gamma^{i}_{jk}$, except four $\Gamma_{00}{}^{0}$, $\Gamma_{11}{}^{1}$, $\Gamma_{22}{}^{2}$, $\Gamma_{33}{}^{3}$, are presented in formulae (41). Connections $\Gamma_{00}{}^{0}$, $\Gamma_{11}{}^{1}$, $\Gamma_{22}{}^{2}$, $\Gamma_{33}{}^{3}$ can be obtained from the last six equations in the general system (51) describing the uniform isotropic matter. Note again, that these equations do not have a unique solution; they can give different solutions, related to the different material fields.

In order to determine from (43) four quantities $\Gamma_{00}^{\ 0}$, $\Gamma_{11}^{\ 1}$, $\Gamma_{22}^{\ 2}$, $\Gamma_{33}^{\ 3}$ (and also twelve other $\Gamma_{jk}^{\ i}$ in (41) depending on them), which are related to some material uniform isotropic field, one can use technique described in [19] for the simple spherically symmetric field.
It means that first of all, we should obtain in U4 space the appropriate equations, tantamount to the equations (43) in G4 space and describing the same uniform isotropic material field. After that, we'll determine the energy-momentum tensor $T_{ik}$ of this material field, then express four connections $\Gamma_{00}^{\ 0}$, $\Gamma_{11}^{\ 1}$, $\Gamma_{22}^{\ 2}$, $\Gamma_{33}^{\ 3}$ in U4 space via tensor $T_{ik}$ components, and finally express four unknown $\Gamma_{00}^{\ 0}$, $\Gamma_{11}^{\ 1}$, $\Gamma_{22}^{\ 2}$, $\Gamma_{33}^{\ 3}$ in G4 space in terms of the $T_{ik}$ components. Two systems of equations (in spaces G4 and U4) must be equivalent, because both provide just different descriptions of one and the same physical field. The reasoning presented in [10], confirms the idea that these two systems are actually equivalent. This equivalency allows determining four unknown connections $\Gamma_{00}^{\ 0}$, $\Gamma_{11}^{\ 1}$, $\Gamma_{22}^{\ 2}$, $\Gamma_{33}^{\ 3}$ in G4 space, if the connections $\Gamma_{00}^{\ 0}$, $\Gamma_{11}^{\ 1}$, $\Gamma_{22}^{\ 2}$, $\Gamma_{33}^{\ 3}$ in U4 space are known. This equivalence might be also viewed as an isometric transition from the affine group to the Lorentz group [17].

System of equations, describing the closed model of a uniform isotropic field in U4 space with matter, should be similar to system (42):

$$R_{00}-\frac{1}{2}g_{00}R=(a')^2+a^2-\frac{1}{3}\Lambda a^4=kT_{00} \quad ,$$

$$R_{11}-\frac{1}{2}g_{11}R=R_{22}-\frac{1}{2}g_{22}R=R_{33}-\frac{1}{2}g_{33}R=2a''a-(a')^2+a^2-\Lambda a^4=kT_{11}=kT_{22}=kT_{33}$$

$$R_{01}=R_{10}=\frac{\partial \Gamma_{11}^{\ 1}}{\partial \eta}-\frac{\partial \Gamma_{00}^{\ 0}}{\partial \chi}=kT_{01}=kT_{10} \quad ,$$

$$R_{02}=R_{20}=\frac{\partial \Gamma_{22}^{\ 2}}{\partial \eta}-\frac{\partial \Gamma_{00}^{\ 0}}{\partial \theta}==kT_{02}=kT_{20} \quad ,$$

$$R_{03}=R_{30}=\frac{\partial \Gamma_{33}^{\ 3}}{\partial \eta}-\frac{\partial \Gamma_{00}^{\ 0}}{\partial \varphi}==kT_{03}=kT_{30}, \tag{64}$$

$$R_{12}=R_{21}=\frac{\partial \Gamma_{22}^{\ 2}}{\partial \chi}-\frac{\partial \Gamma_{11}^{\ 1}}{\partial \theta}==kT_{12}=kT_{21} \quad ,$$

$$R_{13}=R_{31}=\frac{\partial \Gamma_{11}^{\ 1}}{\partial \varphi}-\frac{\partial \Gamma_{33}^{\ 3}}{\partial \chi}==kT_{13}=kT_{31} \quad ,$$

$$R_{23}=R_{32}=\frac{\partial \Gamma_{22}^{\ 2}}{\partial \varphi}-\frac{\partial \Gamma_{33}^{\ 3}}{\partial \theta}==kT_{23}=kT_{32} \quad ,$$

where $T_{ik}$ are the components of the energy-momentum tensor of the respective uniform isotropic material field, $k=8\pi G/c^4=2.07\cdot 10^{-48}$ $s^2/g\cdot cm$ is the Einstein gravitational constant, and c is the speed of light.

System (64) has been obtained using all non-zero components of metric tensor $g_{ii}$ presented in (39) for uniform isotropic space.

The first two equations in (64) can be presented in the closed model as

$$(a')^2 + a^2(1-\frac{k}{3}T_{00}) - \frac{\Lambda}{3}a^4 = 0 \quad ,$$

$$a'' + a\,(1-\frac{k}{6}T_{00} - \frac{k}{2}T_{ii}) + \frac{\Lambda}{3}a^3 = 0 \quad , \qquad (65)$$

where for the uniform isotropic space $T_{ii}=T_{11}=T_{22}=T_{33}$ without summation on indices ii.

For the small $a$, i.e. at the earliest stages of the Universe development during the expansion phase and/or at the latest stages of the Universe development during the contraction phase, when curvature is strong and radius of curvature is small, system (65) in the closed model simplifies:

$$(a')^2 + a^2(1-\frac{k}{3}T_{00}) = 0 \quad ,$$

$$a'' + a\,(1-\frac{k}{6}T_{00} - \frac{k}{2}T_{ii}) = 0 \qquad (66)$$

Respectively, in the open model, acting similarly to what has been done obtaining system (45), it is easy to get the following system of equations:

$$(a')^2 - a^2(1-\frac{k}{3}T_{00}) = 0 \quad ,$$

$$a'' - a\,(1-\frac{k}{6}T_{00} - \frac{k}{2}T_{ii}) = 0 \qquad (67)$$

Emphasize that systems (66)-(67) do not contain terms with cosmological constant $\Lambda$, i.e. according to the physical sense of the cosmological constant, these equations can describe only rather small regions of the uniform isotropic space-time. Also recall that sometimes the existence of the cosmological constant can be considered as the equivalent of the existence of non-zero vacuum energy or existence of the dark energy.

Systems (66)-(67) are similar to systems (49)-(50). Respectively their solutions should be also similar to solutions (51)-(56). The first equations in both systems (66) and (67) are the simple autonomous equations, while the second equations (also in both systems) are the equations of free oscillations. The general solutions of these types of equations are well-known. The problem is to find a solution satisfying both equations simultaneously.

Acting similarly, to what has been done earlier in section 4.4, the general solutions of the first equations in systems (66)-(67) can be presented as

$$a(\eta)=\left|A_1 \exp(\pm i\eta\sqrt{1-\frac{k}{3}T_{00}})\right| \text{ in the closed model },\tag{68}$$

and

$$a(\eta)=\left|A_2 \exp(\pm \eta\sqrt{1-\frac{k}{3}T_{00}})\right| \text{ in the open model,}\tag{69}$$

where $A_1$ and $A_2$ are the constants.

Note that general solutions of the second equations in systems (66) and (67) are similar to the solutions (68)-(69), but have different coefficients in exponents: $\sqrt{1-\frac{k}{6}T_{00}-\frac{k}{2}T_{ii}}$ instead of $\sqrt{1-\frac{k}{3}T_{00}}$ ..

It means the general solutions of systems (66) and (67), satisfying simultaneously both equations in each of these systems, can be obtained only when $T_{00}$=$T_{ii}$=0. It means that uniform isotropic fields, according to the principle of gravitational instability, cannot be a massive one, because any mass increase in some random point will lead to the gravitational attraction to this point, and as result, the field uniformity and isotropy will be ruined. Einstein in [1-2] showed that in $G_4$ space any stationary spherically symmetric material field should be massless, otherwise it cannot be stationary. Therefore, only the massless uniform and isotropic field with $T_{00}$=$T_{11}$=$T_{22}$=$T_{33}$=0 can be really stable at least at the earliest stages of the Universe development during the expansion phase and at the latest stages of the Universe development during the contraction phase.

The initial uniform isotropic state of the Universe, while mathematically possible, is an unstable equilibrium like a perfectly balanced pencil on its tip. This instability is a fundamental mechanism in astrophysics for the formation of cosmic structures like stars, galaxies and clusters overtime from the nearly uniform Universe, which actually was not perfectly uniform and isotropic at least on the small scales even at the early stages of the Universe development during the expansion phase.

Now return to the general system (65), describing the behavior of curvature radius $a(\eta)$ in the closed model of the uniform isotropic field with matter. Acting similarly to what has been done obtaining system (45), we can get the system of equations in the open model:

$$(a')^2-a^2(1-\frac{k}{3}T_{00})-\frac{\Lambda}{3}a^4=0 \quad ,$$

$$a''-a(1-\frac{k}{6}T_{00}-\frac{k}{2}T_{ii})+\frac{\Lambda}{3}a^3=0 \tag{70}$$

The first equations in (65) and (70) are the first order autonomous equations. The general solutions of these equations, similar to solutions (57) and (58), can be presented as

$$a(\eta)=\left|\frac{\pm 2C_1\sqrt{3}(1-\frac{k}{3}T_{00})\exp(\pm i\eta)}{\Lambda C_1^{\ 2}-\exp(\pm 2i\eta)}\right| \quad \text{in the closed model} \tag{71}$$

and

$$a(\eta)=\left|\frac{\pm 2C_3\sqrt{3}(1-\frac{k}{3}T_{00})\exp(\pm \eta)}{\Lambda C_3^{\ 2}-\exp(\pm 2\eta)}\right| \quad \text{in the open model} \tag{72}$$

The second equations in (65) and (69) are the second order autonomous equations. Their general solutions can be obtained as the solutions of the integral equation (48); these solutions have rather complex form and can be expressed via the elliptic integrals of the first kind.

However, the obtained solutions (71)-(72), which are the general solutions of the first equations in (65) and (70), also satisfy the second equations in these systems, but only if $T_{00}=T_{11}=T_{22}=T_{33}=0$, can be proved by substitution. It means that formulae (71)-(72) at $T_{00}=0$, i.e. formulae (57)-(58) represent also the particular solutions of the second equations in systems (65) and (70). Respectively, it means again that the uniform isotropic field should be only massless, because if it is a massive one, it cannot be stable and uniform: see explanation presented above.

However, the obtained simplest particular solutions should have two arbitrary constants. Therefore, we again come to formulae (59)-(60).

Now we can move to the determination of functions $a(t)$ for the closed and open models of the Universe. Using formulae (59) and (40), we can get formulae (61) and determine radius $a(t)$ of the space curvature for the material field in the closed model. Based on the same reasoning, formulae (62) represent radius $a(t)$ of space curvature for the material field in the open model. The discussion of these formulae has already been performed in sections 4.4 and 4.5.

Now let us specify the energy-momentum tensors $T_{ik}$ for a few different uniform isotropic material fields.

Symmetric energy-momentum tensor of the electro-magnetic field can be presented as:

$$T_{ik}=\frac{1}{4\pi}(-F_{il}F_k^l+\frac{1}{4}F_{lm}F^{lm}g_{ik}) \ , \tag{73}$$

where $g_{ik}$ for uniform isotropic field are presented in (39) and $F_{ik}$ is the anti-symmetric tensor of the electro-magnetic field

$$F_{ik}=\frac{\partial A_k}{\partial x_i}-\frac{\partial A_i}{\partial x_k}=\begin{pmatrix} 0 & E_x & E_y & E_z \\ -E_x & 0 & -H_z & H_y \\ -E_y & H_z & 0 & -H_x \\ -E_z & -H_y & H_x & 0 \end{pmatrix}, \qquad (74)$$

where $A_i$ is the 4D potential $A_i$=(φ,-**A**), φ is the scalar potential, **A** is the vector potential, $E_i$ are the components of the electric field intensity, and $H_i$ are the components of the magnetic field intensity.

Symmetric energy-momentum tensor of system of the macroscopic bodies, which are treated as being continuous, can be presented as:

$$T_{ik}=(p+\varepsilon)u_i u_k - p g_{ik}=\begin{pmatrix} \varepsilon & S_x/c & S_y/c & S_z/c \\ S_x/c & -\sigma_{xx} & -\sigma_{xy} & -\sigma_{xz} \\ S_y/c & -\sigma_{yx} & -\sigma_{yy} & -\sigma_{yz} \\ S_z/c & -\sigma_{zx} & -\sigma_{zy} & -\sigma_{zz} \end{pmatrix}, \qquad (75)$$

where p is the pressure, ε is the energy density, $u_i$ are the components of 4D speed, **S** is the vector of energy flux density, and $\sigma_{ik}$ is the momentum flux density.

Symmetric energy-momentum tensor of the scalar field Φ equals:

$$T^{\mu\nu}=g^{\mu\nu}[-\frac{1}{2}\frac{\partial\Phi}{\partial x_\alpha}\frac{\partial\Phi}{\partial x^\alpha}-V(\Phi)]+\frac{\partial\Phi}{\partial x^\mu}\frac{\partial\Phi}{\partial x^\nu}, \qquad (76)$$

where $g_{ik}$ for the uniform isotropic field are determined in (47), and V(Φ) contains all the mass and self-interacting forms of the scalar field Φ.

Symmetric energy-momentum tensor of the Dirac spinor field Ψ is equal to:

$$T^{jn}=\frac{\partial\overline{\Psi}}{\partial x^\mu}\frac{\partial L}{\partial(\frac{\partial\overline{\Psi}}{\partial x^\mu})}+\frac{\partial L}{\partial(\frac{\partial\Psi}{\partial x^\mu})}\frac{\partial\Psi}{\partial x^\nu}-g^{jn}L, \qquad (77)$$

where $\overline{\Psi}$ is the conjugate spinor field and L is the Lagrangian density

$$L=\frac{i}{2}\overline{\Psi}\gamma^\mu\Psi-\frac{i}{2}\frac{\partial\overline{\Psi}}{\partial x^\mu}\gamma^\mu\Psi-m\overline{\Psi}\Psi, \qquad (78)$$

where $\gamma^\mu$ are the Dirac (gamma) matrices, and m is the mass of the spinor particle.

Now using systems of equations (39), (64)-(72) and expressions (73)-(78) for the energy-momentum tensors of different uniform isotropic material fields, it is possible to calculate functions $a(\eta)$ and $\Gamma^i_{jk}(\eta,\chi,\theta)$, i.e. determine the Universe structure in the past, present and future for the different material uniform isotropic fields.

Repeat that initial uniform isotropic state of the Universe is an unstable equilibrium. This instability is a fundamental mechanism in astrophysics for the formation of cosmic structures like stars, galaxies and clusters overtime from the nearly uniform Universe.

Recall again that two systems of equations in $G_4$ and $U_4$ spaces must be equivalent, because both provide descriptions of one and the same physical field. That equivalency allows determining four unknown connections $\Gamma_{ii}^{\ i}$ in (41)-(43) in $G_4$ space, if the connections $\Gamma_{ii}^{\ i}$ in (41) and (64) in $U_4$ space are known. Equations (41) and (64) in $U_4$ space should be tantamount to the equations (41)-(43) in $G_4$ space. The only exceptions are the right-hand sides in equations (64), containing the energy-momentum tensor $T_{ik}$ of the respective material field. Strictly speaking, this tensor in the uniform isotropic space $U_4$ can be calculated by using the respective formulae (73)-(78) for different material fields, but the obtained results will describe, as it has already been mentioned above, the unstable gravitational equilibrium.

Now let us address other important issues. Of course, the obtained results will depend on the values of the used physical constants and parameters characterizing the respective material field. Generically, the structure of the Universe is unstable with respect to the numerical values of the fundamental constants. These constants determine the appearance of the Universe and its stable components: nuclei, atoms, stars and galaxies. Our physical laws and values of the fundamental constants are not only sufficient for the stable conditions of the main Universe components, mentioned above, but they are also necessary; see [29] and references therein.

Comparing the calculated results with the observations, it is possible in principle to solve the inverted problem, i.e. determine the values of physical parameters of the respective material field, at which the calculated model of the Universe is confirmed by the observations. Such concept will be to some extent similar to the cosmological natural selection – hypothesis proposed by Smolin (see e.g. [30] and references therein) and based on the suggestion that a process, analogous to the biological natural selection, applies also at the grandest of the scales – to the Universe.

Recall that generically the inverse problem (or synthesis problem) is the process of calculating from a set of observations the causal factors that produced them, i.e. the unknown parameters that characterize the system.

Note also that all models of Universe are based on the self-referencing physical theories. Therefore, according to the Gödel theorems, one might expect these models to be either inconsistent and/or incomplete, depending on a number of fundamental constants and physical parameters characterizing the respective material field. Values of these constants and parameters cannot be deduced from the first principles, but have to be chosen to fit the observations. It means that all Universe models are incomplete and/or inconsistent [31].

Remind that Gödel theorems prove that any non-trivial system is either incomplete or inconsistent. If system is incomplete, it means there are always will be questions, that cannot be answered using a certain set of axioms, or there are always will be statements that are true, but cannot be proved within the system. If system is inconsistent, it means it is based on the set of axioms, which cannot demonstrate its own consistency, unless a different set of axioms is used.

In order to determine from (43), using condition (12), all components of metric $g_{ik}$ and all non-zero connections $\Gamma^{i}_{jk}$, representing some concrete simple material field, the technique described in [19], where it has been applied to the massless Weyssenhoff fluid with spin [20], can be used.

## 5. Quantization of general equations

Many attempts have been made to quantize equations of the classical geometrical field theories including the GR with the goal to combine quantum mechanics with the GR and/or with different unified geometrical field theories and avoid the contradictions between the quantum and classical theories. At the first glance, it seems that in order to quantize the obtained general equations (5)-(6) we can apply one of the techniques, typically used in attempts to quantize the gravitational or any unified geometrical field: the canonical quantization, path integral quantization, etc. [3-15]. As result, using equations (5)-(6), describing the curvature and torsion of the space-time (i.e. the distribution and motion of matter), we will obtain the quantum equations for "the primary building blocks of everything", i.e. the equations for the "basic (primordial) particles", that form all existing elementary particles and fields; and such "primordial particles" can be represented as some local concentrations of the curvature and torsion of the space-time.

However, there are no rigorous and well-grounded methods used for quantization of the gravitational or any unified geometrical field. All existing methods have serious problems and limitations [3-15], mainly related to the complexity and nonlinearity of the respective equations, entailing violation of the superposition principle, and to the covariance of these theories, leading to the problem with number of the dynamic variables.

The other most important point is that there is probably no need at all to quantize the general equations (5)-(6). If we could quantize equations (5)-(6), then the respective quantum equations describing the "primordial particles" would be obtained. These quantum equations would definitely provide a probabilistic description of the "primordial particles". In quantum mechanics and quantum electrodynamics, a probabilistic behavior of a micro-particle is related to the impossibility to take into account various forces between the specified micro-particle and all other particles. It is obvious, that such interactions always exist and, unlike the macro-objects, they are significant for any micro-particle. In other words, a micro-particle is sensitive even to the very weak interactions; and the number of such interactions is practically unlimited, because a micro-particle "feels" the unified field always existing in any point by interacting with all propagating "waves" coming from everywhere. At the same time, the exact calculation of all such interactions for a micro-particle is impossible in principle.

This reasoning is similar to the de Broglie-Bohm interpretation of quantum mechanics, which is the causal, deterministic, and explicitly nonlocal theory (Bohmian mechanics) [32. Recall that Mach's principle (at least, in one of the interpretations) also means that all local physical phenomena and interactions have the global character, i.e. the parameters of particles, bodies and fields are determined by the global distribution of matter.

Based on the presented ideas, all aspects of a micro-particle behavior (i.e. the equations for the wave function, the Heisenberg uncertainty principle, and many other quantum phenomena) can be described using only the probabilistic quantum methods. The fundamental cause of such probabilistic behavior of a micro-particle is the nonlocal theory describing this particle. In other words, it is just a lack of knowledge that accounts for the uncertainty of many phenomena in quantum mechanics [32].

However, in our case, there is an absolutely different situation concerning the "primordial particles". The equations, determining the curvature and torsion of $G_4$ space in the most general way (or, in other words, determining the distribution and motion of matter consisting of various types of the "primordial particles" and fields), include already all possible interactions among these "primordial particles". Everything (all "primordial particles" and all fields) should be within the scope of such a general approach. This is the "physical sense" of the general equations (5)-(6) or their possible quantum counterparts. Therefore, no probabilistic phenomena and uncertainty should be in the behavior of the "primordial particles". But on the other hand, the probabilistic quantum effects and the respective uncertainty will be definitely presented in the quantized equations.

This problem can be easily solved, if we refuse in principle from the quantization of general equations (5)-(6). It means, the unified geometrical field should not be quantized at all; such field should be described using only the deterministic approach and only classical (not quantum!) equations.

May be, just because of this reason, all the existing theories of quantum gravitational field and quantum unified fields have serious problems and limitations [3-15]. This reasoning confirms the respective ideas of Einstein, Wheeler and others [1-2, 4-5], that only the deterministic classical models, but not the probabilistic quantum ones, should be used to describe various physical fields and particles. For example, Einstein, being one of the creators of quantum theory, of course acknowledged the great achievements of quantum mechanics, but he also saw the principal limitations and "probabilistic issues" of this theory. Therefore, he considered that the elementary particles are just the local concentrations of "strength" of the unified geometrical space-time, and respectively quantum mechanics will be in future just a particular (but very important) part of the general unified field theory. As result, there is no room for a fundamental randomness in the unified geometrical field theory.

Different authors developed similar ideas [7, 10-11, 13], concerning the principal classical character of the gravitational field.

Finally note that results, obtained above in Sections 3-4 using the unified field theory based on the curved and torsional space-time, were used in [19] to provide the hypothetical qualitative explanations of some “non-conventional” phenomena (telepathy, time travel, homeopathy, astrology, and others).

## 6. Conclusions

1. General equations of the unified field theory, obtained using the curved and torsional space-time, are presented. They contain only independent geometrical parameters of the metric-affine space $G_4$: metric $g_{ik}$ and non-symmetric connections $\Gamma_{kl}^{\ \ i}$. As result, these equations describe the distribution and motion of matter, which represents the curvature and torsion of the space-time unified geometrical field.
2. Various types of curvature and torsion in $G_4$ space are related to different physical fields. The simplest set of the non-symmetric affine connections $\Gamma_{kl}^{\ \ i}$ describes the space-time with minimum curvature and torsion. Such “simplest” physical field is a pure gravitational field in vacuum. The more complex sets of connections $\Gamma_{kl}^{\ \ i}$ represent the more complex geometries of the space-time and describe the more complex material physical fields.
3. Solutions of the general equations, describing spherically symmetric fields, were obtained for gravitational field in vacuum and arbitrary material field. The obtained results endorse the concept that metric-affine space $G_4$ with curvature and torsion describes the distribution and motion of matter. They also confirm correctness of the general equations and accuracy of their spherically symmetric solutions for gravitational field in vacuum, because these solutions are stationary, do not contain singularities, and in the limit coincide with the Schwarzschild solution.
4. Solutions of the general equations, describing the uniform isotropic fields with and without cosmological terms for the closed and open models of the Universe, were obtained for gravitational field in vacuum and arbitrary material field. The obtained results confirm validity of the general and approximate equations, and correctness of the obtained solutions, related to the physical fields with minimum curvature and torsion – the uniform isotropic gravitational fields in vacuum, because these solutions do not contain singularities and are similar (at least, qualitatively) to the respective solutions describing other models of the Universe: cyclic closed model, Friedmann model, de Sitter model, Big Bounce open model, and “dust-like” matter in the closed model.
5. Equations, describing cosmological models, are presented for various material uniform isotropic fields, but their solutions (formulae for the curvature radius) have been obtained only for the simplest cases – the massless uniform isotropic fields. The relations of the obtained results to the anthropic principle, cosmological natural selection and Gödel theorems, are discussed.
6. It is shown that there is no necessity to quantize the obtained general equations, because these equations describe the evolution of matter in the most general way by taking into account all possible interactions between different particles and fields. Therefore these equations exclude probabilistic phenomena and uncertainty, and show that any physical field (i.e. the respective type of the curved and torsional space-time) can be described using only the deterministic approach.